\documentclass[12pt,preprint]{aastex}
\usepackage{natbib}

\shorttitle{LMXBs and GCs in Early-Type Galaxies}
\shortauthors{SARAZIN ET AL.}
\slugcomment{Astrophysical Journal, submitted}

\begin{document}

\title{Low Mass X-ray Binaries and Globular Clusters in Early-Type Galaxies}

\author{Craig L. Sarazin\altaffilmark{1},
Arunav Kundu\altaffilmark{2},
Jimmy A. Irwin\altaffilmark{3,4},
Gregory R. Sivakoff\altaffilmark{1}, \\
Elizabeth L. Blanton\altaffilmark{1,4},
and
Scott W. Randall\altaffilmark{1}
}

\altaffiltext{1}{Department of Astronomy, University of Virginia,
P. O. Box 3818, Charlottesville, VA 22903-0818;
sarazin@virginia.edu,
grs8g@virginia.edu,
eblanton@virginia.edu,
swr3p@virginia.edu,
}

\altaffiltext{2}{Department of Physics and Astronomy, Michigan State
University, East Lansing, MI 48824; akundu@pa.msu.edu}

\altaffiltext{3}{Department of Astronomy, University of Michigan,
Ann Arbor, MI 48109-1090; jirwin@astro.lsa.umich.edu}

\altaffiltext{4}{Chandra Fellow}

\begin{abstract}
A high fraction of the Low Mass X-ray Binaries (LMXBs) in early-type galaxies
are associated with globular clusters (GCs).
Here, we discuss the correlations between LMXBs and GCs in a sample of
four early-type galaxies with X-ray source lists determined from
{\it Chandra} observations.
There is some evidence that the fraction of LMXBs associated with globular
clusters ($f_{\rm X-GC}$) increases along the Hubble
sequence from spiral bulges (or spheroids) to S0s to Es to cDs.
On the other hand, the fraction of globular clusters which contain X-ray
sources appears to be roughly constant at $f_{\rm GC-X} \sim 4$\%.
There is a strong tendency for the X-ray sources to be associated
with the optically more luminous GCs.
However, this correlation is consistent with a constant probability of
finding a LMXB per unit optical luminosity;
that is, it seems to result primarily from the larger number of stars
in optically luminous GCs.
The probability of finding a bright LMXB per unit optical luminosity
in the GCs is about
$1.5 \times 10^{-7}$ LMXBs per $L_{\odot,I}$ for
$L_X \ga 1 \times 10^{38}$ erg s$^{-1}$ (0.3--10 keV)
and rises to about
$2.0 \times 10^{-7}$ LMXBs per $L_{\odot,I}$ at lower X-ray luminosities,
$L_X \ga 3 \times 10^{37}$ erg s$^{-1}$.
This frequency appears to be roughly constant for different galaxies,
including the bulges of the Milky Way and M31.
There is a tendency for the X-ray sources to be found preferentially
in redder GCs, which is independent of optical luminosity correlation.
This seems to indicate that the evolution of X-ray binaries in a GC is
affected either by the metallicity or age of the GC, with younger and/or
more metal rich GCs having more LMXBs.
There is no strong difference in the X-ray luminosities of GC and non-GC
LMXBs.
There is a weak tendency for the brightest LMXBs, whose luminosities
exceed the Eddington luminosity for a 1.4 $M_\odot$ neutron star, to
avoid GCs.
That may indicate that black hole X-ray binaries are somewhat less
likely to be found in GCs,
as seems to be true in our Galaxy.
On the other hand, there are some luminous LMXBs associated with GCs.
There is no clear evidence that the X-ray spectra or variability of GC and
non-GC X-ray sources differ.
We also find no evidence for a difference in the spatial distribution of
GC and non-GC LMXBs.
Many of these results are similar to those found in NGC~1399 and
NGC~4472 by
Angelini et al.\ and
Kundu et al.,
respectively.
\end{abstract}


\section{Introduction} \label{sec:intro}

X-ray observations since the time of the {\it Einstein} Observatory have
shown that early-type galaxies are often luminous X-ray sources
\citep*[e.g.,][]{fjt85}.
For the X-ray-luminous early-type galaxies
(defined as those having a relatively high ratio of X-ray to optical luminosity
$L_X/L_B$),
it is clear that the bulk of the X-ray luminosity is from
hot ($\sim$$10^7$ K) interstellar gas.
On the other hand, recent {\it Chandra} observations have resolved most
of the X-ray emission in the X-ray-faint early-type galaxies into
individual point-like sources
\citep*[e.g.,][]{sib00,sib01}.
Given their properties and the stellar populations in these galaxies,
these X-ray sources are assumed to be Low Mass X-ray Binaries (LMXBs).
LMXBs are also observed in {\it Chandra} observations of
X-ray-luminous early-type galaxies, and often dominate the X-ray emission
at hard X-ray energies
\citep*[e.g.,][]{kra+00, alm01}.

{\it Chandra} observations of early-type galaxies have shown that
a significant fraction ($\sim$20-70\%) of the LMXBs are associated with
globular clusters in the host galaxies
\citep{sib00,sib01,alm01,kmz02}.
The fraction of LMXBs located in GCs is much higher than the
fraction of optical light, which indicates that nondegenerate stars 
in GCs are much more likely (by a factor of $\sim$300) to be donor
stars in X-ray binaries than field stars.
As has been known for a number of years,
a similar result applies to our own Galaxy and to the bulge of M31
\citep{kat75,hg83,wnp95}.
This is generally believed to result from stellar dynamical interactions
in globular clusters, which can produce compact binary systems
\citep*{cla75,fpr75,hil76}.

X-ray observations with {\it ASCA} indicate that the total luminosity of
LMXBs in early-type galaxies correlates better with the number of
GCs than with the optical luminosity of the galaxy
\citep*{wsk02}.
The correlation is consistent with the total X-ray luminosity of LMXBs
being proportional to the number of GCs.
This is somewhat surprising, as a nontrivial fraction ($\sim50$\%) of the
LMXBs in most of the early-type galaxies observed so far with {\it Chandra}
are not identified with GCs.
This suggests that most (perhaps all?) of the LMXBs in early-type galaxies
were made in GCs
\citep{gri84,sib01,wsk02}.
The field LMXBs might have been ejected from globular clusters by
kick velocities resulting from supernovae, by stellar dynamical processes,
or by the dissolution of the globular due to tidal effects.
If this is true, then the field LMXBs are a useful record of the total
globular cluster populations and of the history of globular cluster
destruction in galaxies.

Alternatively, if field LMXBs are not generally made in GCs, then one might
expect field LMXBs and GC LMXBs to have different properties.
Presumably, the stars in field LMXBs are formed in binaries, and evolve
independently (i.e., they are not affected by their environment).
As evidenced by the high X-ray to optical ratio of GCs, LMXBs in GCs
are strongly affected by dynamical interactions within the GC.
For example, it has been argued that it is difficult for GCs to retain
LMXBs with massive black holes (BHs), which may be ejected by dynamical
interactions
\citep[e.g.,][]{pzm00},
which is consistent with the fact that there are presently no BH LMXBs
identified with GCs in our own Galaxy.
Thus, one might expect very luminous LMXBs to avoid GCs in early-type
galaxies.

One might also expect LMXBs in different GCs to be affected by differences
in their GC environment.
This might occur because of differences in the ages or metallicities of
different GCs, which might affect the evolution of individual LMXBs.
Alternatively, differences in stellar densities would change the rates
of stellar dynamical interactions.

In this paper, we give a preliminary analysis of the correlations of
GC LMXBs and non-GC LMXBs, and of the differences between X-ray GCs
and non-X-ray GCs in early-type galaxies.
This analysis should be viewed as preliminary because of the small size
and incompleteness of the existing samples.
At present, only a relatively small number of early-type galaxies have been
observed with {\it Chandra}.
The spatial resolution of {\it Chandra} is needed to separate LMXBs at
the distance of the Virgo cluster, where most of the nearby early-type
galaxies are located.
Moreover, most of the galaxies observed with {\it Chandra} so far are
X-ray bright galaxies.
In these galaxies, bright diffuse emission from hot gas makes it difficult
to detect LMXBs near the center of the galaxy
\citep[e.g.,][]{alm01}.
The optical samples of GCs also are incomplete.
Ground-based observations have difficulty in detecting GCs near the centers
of early-type galaxies, where they are blended into the diffuse optical
emission from the galaxy.
Thus, ground-based GC samples tend to be restricted to the outer parts
of the galaxies, which is not where most of the LMXBs are located.
Also, ground-based observations do not generally resolve GCs at the
distance of the Virgo cluster.
Thus, there is a significant chance that ground-based GCs detected in
the outer parts of early-type galaxies may actually be foreground or
background objects.
GC observations with the {\it Hubble Space Telescope} ({\it HST})
are greatly preferable, as the resolution of {\it HST} allows GCs to be
detected in the inner regions of galaxies, and slightly resolves them.
However, some of the galaxies with the best X-ray data have no
{\it HST} GC observations (e.g., NGC~4697).
Also, the field of view of a single Wide Field Planetary Camera 2
(WFPC2) observation is smaller than the angular size of early-type
galaxies at the distance of the Virgo cluster.
Multiple pointings with the WFPC2
\citep[e.g.,][]{kmz02}
or with the recently installed Advanced Camera for Surveys
(ACS) may improve this situation.

Because the GC observations of most galaxies do not cover the entire
galaxy, we will restrict our analysis to the LMXBs located in the regions
covered by the GC sample.
We will also restrict our discussion to the relative properties of
GC LMXBs and non-GC LMXBs, and of X-ray GCs and non-X-ray GCs.
Hopefully, comparing the relative properties of these samples within
the spatial regions they share
should reduce most of the selection effects in these incomplete samples,
particularly as the LMXBs were detected solely by their X-ray properties
and the GC were detected solely by their optical properties.
All other things being equal, there is no reason why a LMXB in
a GC is any more or less detectable in the X-ray than a field LMXB.
Because of the irregular spatial regions covered by the GC samples,
we will concentrate on the intrinsic properties of the sources rather
than their spatial distribution.

Our galaxy sample, X-ray source lists, and the detection limits for
the LMXBs are discussed in \S~\ref{sec:sample}.
The statistics of occurrence of LMXBs in the galaxies are given
in \S~\ref{sec:statistics}.
The relative properties of GC and non-GC LMXBs are analyzed in
\S~\ref{sec:gc}, while the relative characteristics of
X-ray and non-X-ray GCs are determined in \S~\ref{sec:xray}.
In \S~\ref{sec:spatial},
we briefly discuss the relative spatial distributions
of the samples.
We discuss our results and summarize our 
conclusions in \S~\ref{sec:conclusion}.

\section{Sample} \label{sec:sample}

Table~\ref{tab:galaxies} lists some of the properties of the
early-type galaxies in our sample.
The first five columns give
the name,
Hubble type,
Galactic absorbing column $N_H$ \citep{dl90},
effective radius $r_{\rm eff}$,
and
distance $D$.
For consistency, all of the distances for these galaxies are taken from
\citet{tdb+01},
based on the method of surface brightness fluctuations.
For these galaxies, the statistical error in their distance modulus
is generally about 0.17, leading to an error in the distance of
8\%.
There also is a systematic error (which would affect all of the
galaxies by the same factor) of about 8\% in the distance.

The last 5 columns in Table~\ref{tab:galaxies} give
the useful exposure time in the {\it Chandra} X-ray image,
the minimum detectable source count rate,
and the minimum detectable X-ray luminosity in the 0.3--10 keV band,
and the references for the original papers giving the detections of
the LMXBs and the GCs.
The criteria for the detection of the X-ray sources are discussed
in detail in our original papers on the X-ray observations.
These papers also list the properties, including the count rates, for each
of the sources.
In every case, a wavelet detection algorithm
({\sc ciao wavdetect}\footnote{See \url{http://asc.harvard.edu/ciao/}.}
program),
was used to detect the sources, with a significance threshold of
$10^{-6}$.
This implies that $\la$1 false source
(due to a statistical fluctuation in the background) would be detected
in the entire S3 image.
We further restricted the sources by requiring
sufficient source counts to determine the source flux at the $\ge 3 \sigma$
level.
In general, this implies that the minimum detectable sources had $\ga$10
net counts.

For the Virgo cluster galaxies, some of the X-ray sources have
been identified with outer globular clusters from the lists given by
\citet{han77}.
In general, less information is available for these outer globulars.
Also, there is a higher chance that the globulars and X-ray sources at
large distances from the centers of the galaxies are actually foreground
or background objects.
Thus, we haven't included these globulars and the associated X-ray sources
in our sample.

Table~\ref{tab:src} lists some of properties of the X-ray sources in
the region covered by the GC surveys.
More detailed information is given in the original references on the X-ray
detections, which are listed in Table~\ref{tab:galaxies}.
Table~\ref{tab:src} gives the galaxy,
the source number from the tables in the original references,
the net X-ray count rate and error in the 0.3--10 keV band,
the X-ray luminosity $L_X$ in the same band (\S~\ref{sec:xray_lum} below),
the X-ray hardness ratios $H21^0$ and $H31^0$
(\S~\ref{sec:xray_colors} below),
an indication of whether the source was found to be variable
(\S~\ref{sec:xray_vary} below),
a check if the source was identified with a GC,
the I-band absolute magnitude $M_I$ of the GC (\S~\ref{sec:gc_mag} below),
and the $V-I$ color of the GC (\S~\ref{sec:gc_color} below).
The last column has a check if the source belongs to an X-ray luminosity
limited complete sample which is defined in \S~\ref{sec:sample_combined}
below.

Figure~\ref{fig:regions} shows the regions surveyed to detect GCs
in these galaxies.
These are compared to the circle with a radius of one effective radius.

\subsection{NGC~1553} \label{sec:sample_n1553}

The globular cluster identifications in NGC~1553 come from a survey
of globulars in S0 galaxies with the {\it HST} Wide Field / Planetary
Camera 2
\citep[WFPC2;][]{kw01b}.
This GC sample included the nucleus of the galaxy and the inner parts of
the galaxy to the southeast of the nucleus, out to a radius of about
2\arcmin\ 
(Fig.~\ref{fig:regions}).

The X-ray sources were published in \citet{bsi01}.
We adjusted the values of the X-ray luminosities to be consistent with
the new distance to this galaxy
\citep{tdb+01}.
We excluded the central X-ray source (Src.~1, CXOU J041610.5$-$554646)
because it is the central active galactic nucleus (AGN).
Srcs.~2 (CXOU J041609.9$-$554646) and 6 (CXOU J041611.7$-$554652)
were excluded because they are located near the center of the galaxy
and 1-2\arcsec\ from several globular cluster positions.
This is too large a separation for a definite identification in this
crowded region, so we were uncertain whether to associate these sources
with the GCs or not.

The globular cluster identifications are discussed in \citet{bsi01}.
We used the position of the central AGN, which is present in both the
{\it HST} optical and {\it Chandra} X-ray images, to register the two
images, assuming a simple shift.
This required an offset of 1\farcs9.
[We also registered the images individually by comparison to accurate
optical positions from 
the U.S. Naval Observatory A2.0 optical catalog
\citep[USNOA2][]{Mon+98} and
the Two Micron All Sky Survey
\citep[2MASS][]{Cut+01}.
This indicated that both the {\it Chandra} and {\it HST} positions were
off by $\sim$1\farcs5, and that the relative offset was the combination
of these systematic errors.]
Three potential GCs were found to agree in position with LMXBs to within
0\farcs5.
Based on simulations of the distributions of GCs and LMXBs, the expected
number of random associations would be about 0.05.
However, one of the potential GCs had a very blue color, and was dropped
from the sample using the criteria for GC identifications given in
\citet{kw01b}.

\subsection{NGC~4365} \label{sec:sample_n4365}

\citet*{ssi03} have determined the X-ray source population in NGC~4365.
The globular cluster population is from an {\it HST} WFPC2 survey of elliptical
galaxies by
\citet{kw01a}.
This GC sample included the nucleus of the galaxy and the inner parts of
the galaxy to the northwest of the nucleus, out to a radius of about
2\arcmin\ 
(Fig.~\ref{fig:regions}).
We found an offset of about 1\farcs3 between the {\it HST} and
{\it Chandra} positions.
After removing this offset, 18 X-ray sources were found to lie within
1\farcs02 of a GC.
Simulations of the distributions of GCs and LMXBs indicate that
about 0.6 associations would be expected at random.
One additional association was found between a LMXB and a GC in the short
list of clusters with spectroscopy recently published by
\citet{LBB+03}.
However, it is outside of the region covered by the GC survey in
\citet{kw01a}, and we haven't included this source in
the present study.

\subsection{NGC~4649} \label{sec:sample_n4649}

The X-ray sources in NGC~4649 (M60) have been published in
\citet*{rsi03}.
As with NGC~4365, the globular cluster sample here is from
\citet{kw01a}.
This sample included the nucleus of the galaxy and the inner parts of
the galaxy to the northwest of the nucleus, out to a radius of about
2\arcmin\ 
(Fig.~\ref{fig:regions}).
At the northwest corner of the WFPC2 field, there is a high density of
globular clusters which are probably associated with the companion
Sc galaxy NGC~4647.
Because we wish to study the LMXB population in early-type galaxies, we
have excluded this region from our sample.
We found an offset of 0\farcs9 between the {\it HST} and
{\it Chandra} positions.
After correcting for this offset, 22 GCs associated with NGC~4649
were found to coincide with X-ray sources to better than 0\farcs87.
Based on simulations of the distributions of GCs and LMXBs, the expected
number of random associations would be about 2.96, so a few of these
identification may be spurious.
NGC~4649 is an X-ray bright elliptical galaxy (the only one in our
sample).
Unfortunately, the bright X-ray emission by interstellar gas makes
it difficult to detect faint sources near the center of the galaxy,
and the LMXB source list is incomplete within approximately 70\arcsec\ of
the center of NGC~4649.
This X-ray incompleteness may reduce the detected fraction of GCs
containing X-ray sources, 
$f_{\rm GC-X}$.
Ideally, one would exclude this region from our sample, but this would
leave almost no area of the galaxy which was on the WFPC2 image,
beyond 70\arcsec, and not confused with the companion NGC~4647.
Since our primary aim is a comparative study of globular cluster and
non-globular cluster LMXBs, we have included all of the covered area
of the galaxy.
We exclude X-ray Src.~1 (CXOU J124340.0+113311), which is an extended feature
due to structure in the diffuse gaseous emission.

\subsection{NGC~4697} \label{sec:sample_n4697}

The X-ray source population of NGC~4697 has been analyzed and listed
in
\citet{sib00,sib01}.
We adjusted the values of the X-ray luminosities to be consistent with
the new distance to this galaxy
\citep{tdb+01}.
The X-ray observation was done in {\it Chandra} Cycle-1 at a time
when significant absolute errors in positions occurred.
Thus,
the X-ray image was registered by comparing the positions of X-ray
sources with optical identifications (not the GC discussed below)
to optical positions from the USNOA2 catalog
\citep{Mon+98}.
This required an offset of about 2\farcs6.
Following this correction, the X-ray and optical positions agreed to
better than 0\farcs3.
Subsequently, 2MASS \citep{Cut+01} positions became available for several
X-ray sources with optical IDs, and these confirmed the offset and
corrected accuracy of the X-ray positions.

The globular cluster sample for NGC~4697 comes from an unpublished,
ground-based ccd survey
very kindly provided by J. Kavelaars (2000, private communication).
These globulars are all located in a circular annular region extending
from about 1\farcm5 to 2\farcm5 from the center of NGC~4697
(Fig.~\ref{fig:regions}).
Seven of the X-ray sources were found to lie within 0\farcs78 of
X-ray sources.
From simulations based on the distribution of LMXBs and GCs, we
would expect 0.18 associations to occur at random.
Although the GCs are not resolved in these ground-based observations,
this sample doesn't extend to a large enough projected distance that
an important fraction of the GCs are likely to actually be background
AGN or foreground stars.
It is difficult to detect globular clusters in the central regions of
early-type galaxies with ground-based observations because of the
effects of seeing and the high surface brightness of the galaxy.
The available data on NGC~4697 do not include colors for the globular
clusters, so these GCs will not be used for looking for trends with
color.
Also, the magnitudes for the NGC~4697 GCs were determined in the R-band,
whereas the three other galaxies used the V and I bands.
For comparison, we converted the R magnitudes to I magnitudes assuming that
$I = R - 0.49$, which is appropriate for the average color of the GCs in
the other galaxies.
We also increased the errors to include a systematic error on this conversion.

\subsection{Combined Samples} \label{sec:sample_combined}

In this paper, we will discuss some of the properties of the LMXBs and
associated GCs in the individual galaxies.
However, our primary aim is to compare LMXBs and GCs in the sample formed
by combining the galaxies.
We will consider two combined samples.
First, we will consider an X-ray luminosity limited sample (``$L_X$
complete'') of sources whose X-ray luminosities exceed
$1.23 \times 10^{38}$ erg s$^{-1}$ in the 0.3--10 keV band, which is the
highest detection limit for the galaxies in the sample
(Table~\ref{tab:galaxies}).
In order to improve the statistics,
we will also consider the sample formed by combining all of the sources,
although this sample is not complete in the X-ray luminosity,
However,
we will restrict our discussion to the relative properties of
GC LMXBs and non-GC LMXBs, and of X-ray GCs and non-X-ray GCs.
Hopefully, comparing the relative properties of these samples within
the spatial regions they share
should reduce most of the selection effects in this incomplete sample,
particularly as the LMXBs were detected solely by their X-ray properties
and the GC were detected solely by their optical properties.
All other things being equal, there is no reason why a LMXB in
a GC is any more or less detectable in the X-ray than a field LMXB.

\section{Statistics of Occurrence of LMXBs} \label{sec:statistics}

The numbers of LMXBs and GCs in our sample are listed in
Table~\ref{tab:statistics}.
The first three columns give
the galaxy name,
the Hubble type,
and 
the number of GCs in the surveyed region $N_{\rm GC}$.
The next five columns give
the number of LMXBs in the same region $N_{\rm LMXB}$,
the number of LMXBs coincident with GCs $N_{\rm both}$,
the fraction of LMXBs identified with GCs
$f_{\rm X-GC} \equiv ( N_{\rm both}/ N_{\rm LMXB} ) $,
the fraction of GCs associated with LMXBs
$f_{\rm GC-X} \equiv ( N_{\rm both}/ N_{\rm GC} ) $,
and
the number of LMXBs in GCs divided by their total optical luminosity.
The LMXBs included in these columns are only those in the $L_X$ complete
sample
($L_X > 1.23 \times 10^{38}$ erg s$^{-1}$ in the 0.3--10 keV band).
The errors listed are 1--$\sigma$;
errors on fractions are derived assuming a binomial distribution.
The next five columns repeat the statistics for the full sample of LMXBs.

Note that the elliptical galaxies in our sample all have a similar fraction
of LMXBs in GCs ($f_{\rm X-GC} \approx 50$\%).
The value for NGC~4697 in the complete sample is larger, but the errors
are very big because of the small number of sources.
In any case, the hypothesis that the number of LMXBs in GCs in NGC~4697
was drawn from the same distribution as that of the other ellipticals
can only be rejected at the 57\% confidence level.
In the full sample, the fraction in NGC~4697 is in good agreement
with those of the other ellipticals.
A similar fraction was found for the X-ray bright elliptical NGC~4472
by \citet{kmz02}.

Although the statistics are very poor, the fraction in the one S0 galaxy
in our sample (NGC~1553) is lower, $f_{\rm X-GC} = 18$\%.
We tested the hypothesis that the number of LMXBs in GCs in NGC~1553
was drawn from the same distribution as that for the elliptical galaxies,
and found that it could be rejected at the 99\% confidence level.
The fraction in spiral bulges or spheroids appears to be even lower,
$f_{\rm X-GC} \sim 10\%$ \citep[e.g.,][]{vvdh95}.
For example, in our Galaxy there are 14 LMXBs which have been associated
with GCs
\citep{wnp95,har96,lvv01,wa01} out of a total of about 150 LMXBs.
This would suggest that $f_{\rm X-GC} \approx 9$\%.
However, most of the LMXBs seen in our Galaxy are much too faint to have
been detected in distant galaxies.
In the Milky Way, there are very few LMXBs with persistent X-ray
luminosities as high as those for our $L_X$ complete sample.
So, we instead compare to the fraction seen in NGC~4697, where the
luminosity limit is $L_X > 2.7 \times 10^{37}$ erg s$^{-1}$.
If we limit the sample to sources which are more luminous than this
for periods of many hours (i.e., not the burst luminosity), then the fraction
is very roughly $f_{\rm X-GC} \approx 7$\%.
At similar luminosity levels, the fraction of bulge or spheroid X-ray sources
in M31 which are identified with GCs is about 20\%
\citep*{pfj93,shp+97,bh01,dkg+02}.
On the other hand,
\citet{alm01} report a fraction of $\sim$70\% in NGC~1399, the cD galaxy at
the center of the Fornax cluster.
Although the number of galaxies tested so far is small and the statistics
in most of the galaxies are poor,
all of this is consistent with an increase in $f_{\rm X-GC}$ along the Hubble
sequence from spiral bulges to S0s to Es to cDs.
The specific frequency of globular clusters (the number per unit galaxy
luminosity) also increases along the same sequence
\citep[e.g.,][]{har91}.

There is no clear trend in the fraction of GCs containing X-ray sources
($f_{\rm GC-X}$) with Hubble-type.
All of our galaxies are roughly consistent with about 3.3\% of the GCs hosting
luminous LMXBs at the luminosity level of the $L_X$ complete sample
($L_X > 1.23 \times 10^{38}$ erg s$^{-1}$ in the 0.3--10 keV band).
At lower luminosity levels,  the fraction appears to increase to about 4\%.
Similar numbers are found for the spiral bulge or spheroid of our Galaxy and M31
\citep[e.g.,][]{har96,bh01,lvv01,dkg+02}.
If we compare to the fraction seen in NGC~4697, where the
luminosity limit is $L_X > 2.7 \times 10^{37}$ erg s$^{-1}$,
we find that roughly $f_{\rm GC-X} \approx 3\%$.
Since many of the LMXBs are highly variable, and catalogs for our Galaxy
include sources detected at different times, it is not clear how many of
these sources could be detected in a single $\approx 10$ hour observation
with {\it Chandra} if our Galaxy were at the distance of the Virgo cluster.
In the bulge of M31, the fraction of GCs containing LMXBs with $L_X \ga 2.7
\times 10^{37}$ erg s$^{-1}$ is about 2\%
\citep[e.g.,][]{bh01,dkg+02}.
Similar fractions have been found for
the elliptical galaxy NGC~4472 \citep{kmz02},
and for the galaxy NGC~1399 at the center of the Fornax cluster
\citep{alm01}.
In fact, \citet{kmz02} argued that $f_{\rm GC-X}$ was constant at about
4\%\ in a wide range of galaxies.
This fraction may provide a constraint on the duty cycle of quiescence vs.\
X-ray brightness in the LMXBs in GCs.

One concern with the fraction $f_{\rm GC-X}$ is that it may be affected
by incompleteness in the detection of faint GCs.
Luminosity functions of globular clusters in early-type galaxies
are often well fit by log-normal functions
\citep[e.g.,][]{har91}; 
for such a luminosity distribution, faint GC may contribute to the
number of GCs, but do not add significantly to the total luminosity.
Below, we will find that LMXBs are preferentially associated with
more luminous GCs, but that the probability of finding a LMXB in a
GC is roughly proportional to its optical luminosity.
Thus, a more useful measure of the abundance of LMXBs in GCs may be
the number per unit optical luminosity, which should be less sensitive to
incompleteness or to the details of the luminosity function of GCs in
each galaxy.
This is given as the last numerical column in Table~\ref{tab:galaxies}.
For the entire sample of four galaxies, the total value is
$1.5 \times 10^{-7}$ LMXBs per $L_{\odot,I}$ in the $L_X$ complete sample
and $2.0 \times 10^{-7}$ LMXBs per $L_{\odot,I}$ in the full sample.
To within the errors, all of the galaxies are consistent with this
value.
At luminosities $L_X \ga 2.7 \times 10^{37}$ erg s$^{-1}$ (2--10 keV), the
corresponding value for our Galaxy is about
$1.4 \times 10^{-7}$ LMXBs per $L_{\odot,I}$.

\section{GC Properties of LMXBs} \label{sec:gc}

\subsection{Luminosities} \label{sec:gc_mag}

\begin{figure}
\plottwo{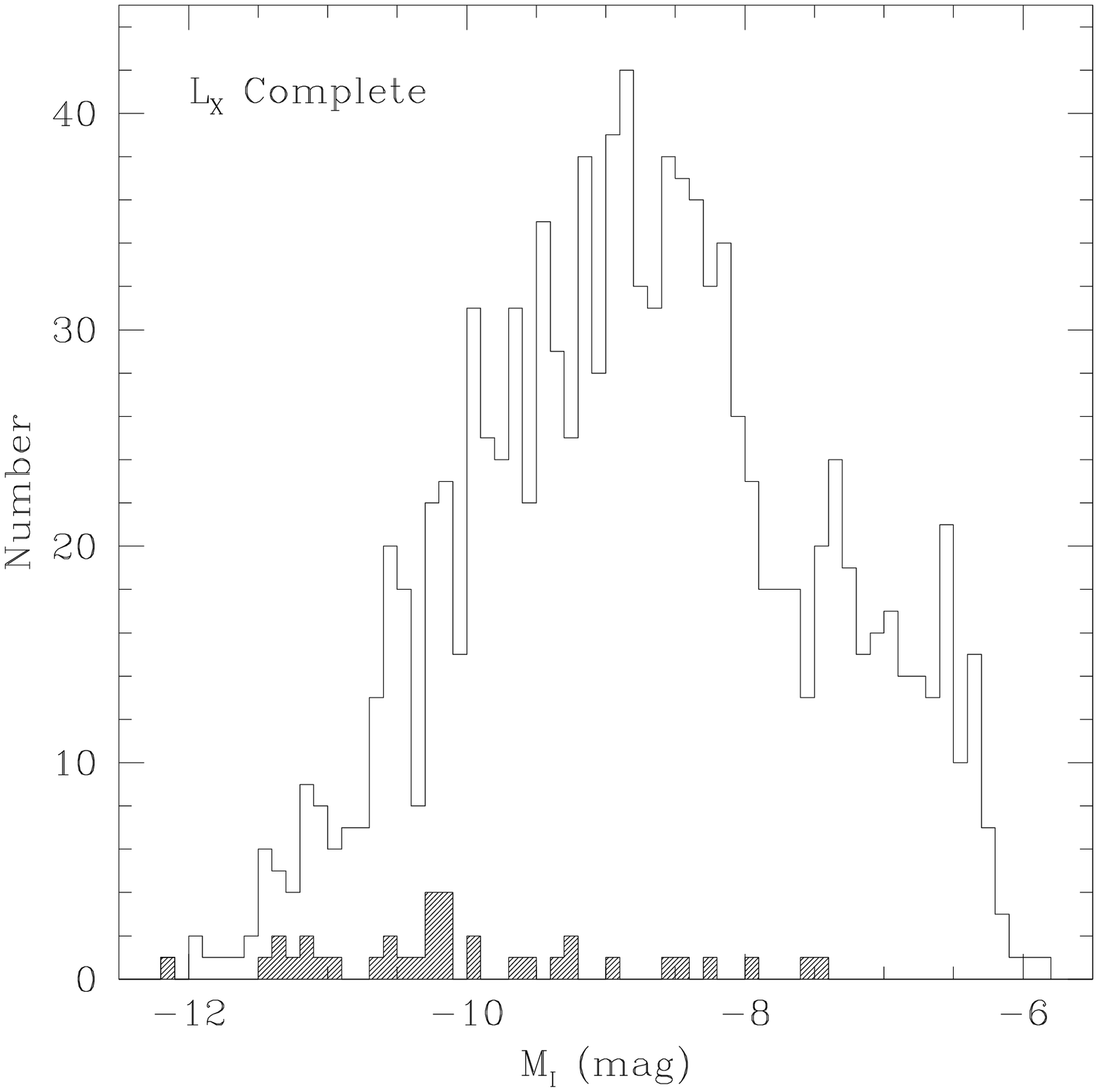}{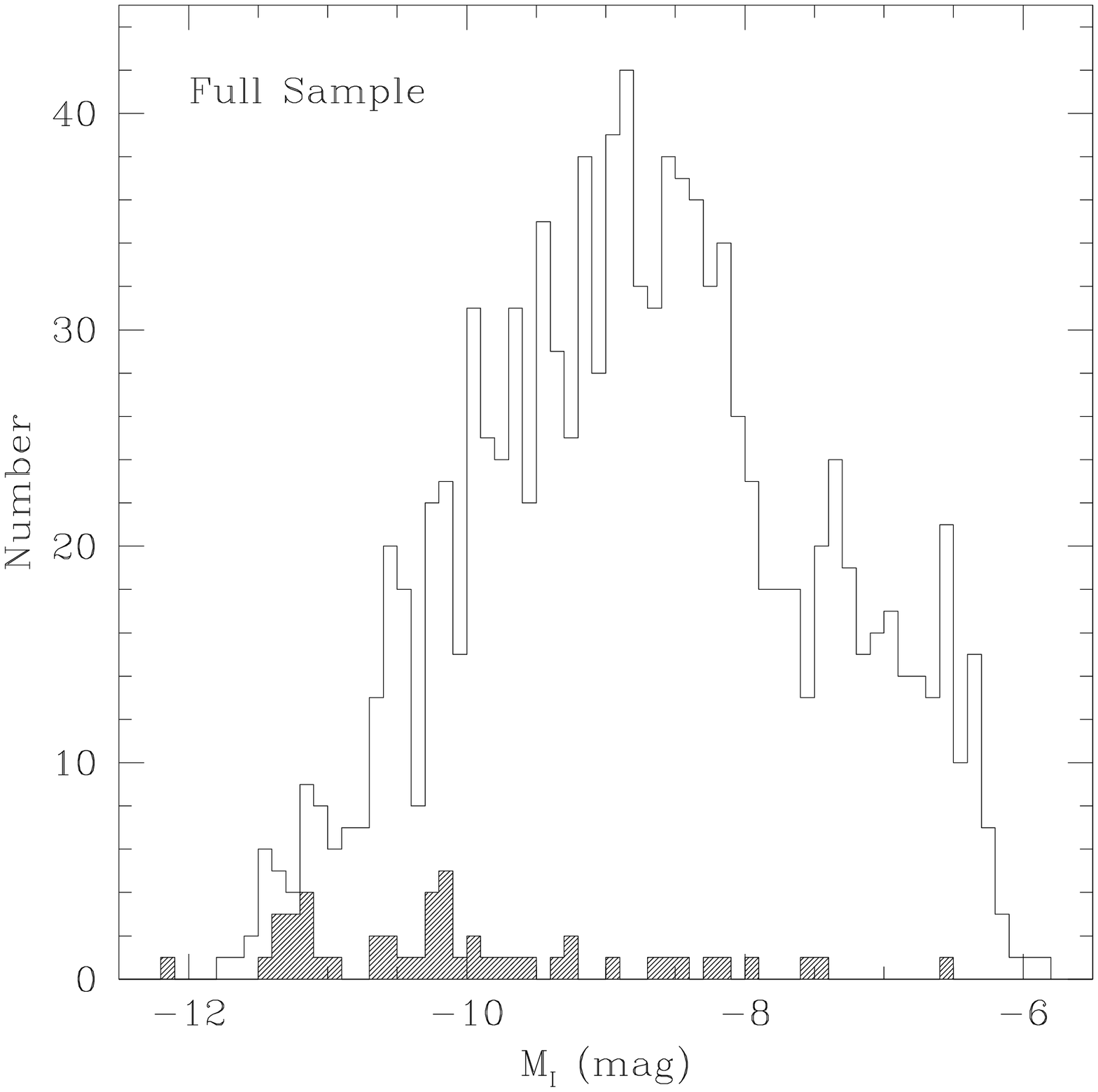}
\caption{Histograms of the number of globular clusters versus their
absolute magnitude, $M_I$.
The left panel is for the $L_X$ complete sample, while the right panel is
for the full sample.
In each case,
the upper histogram is for all of the GCs in the galaxies.
The lower shaded histogram shows the GCs which contain identified LMXBs.
The histogram bins are 0.1 mag wide.
\label{fig:gc_mag}}
\end{figure}

\begin{figure}
\plottwo{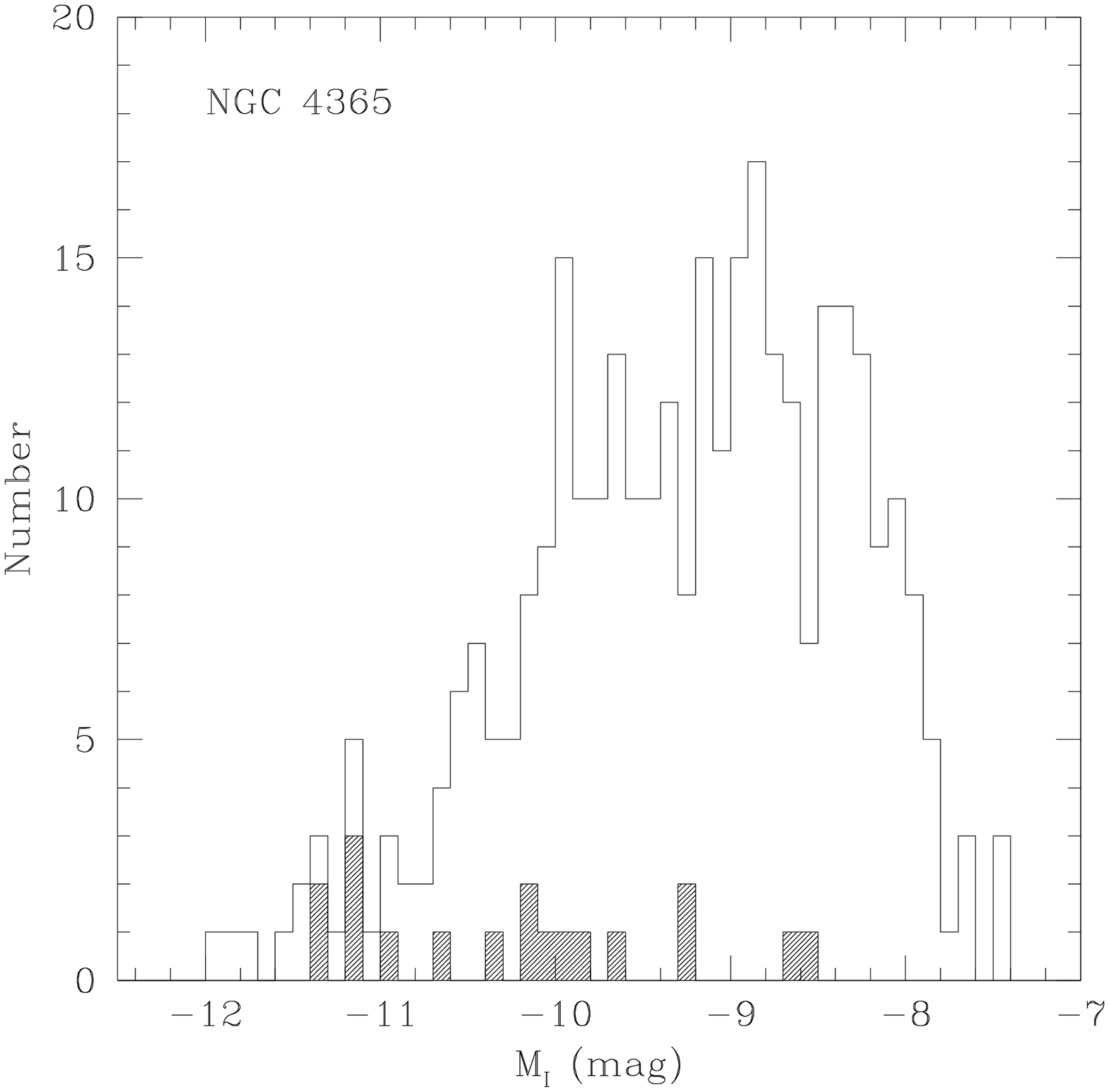}{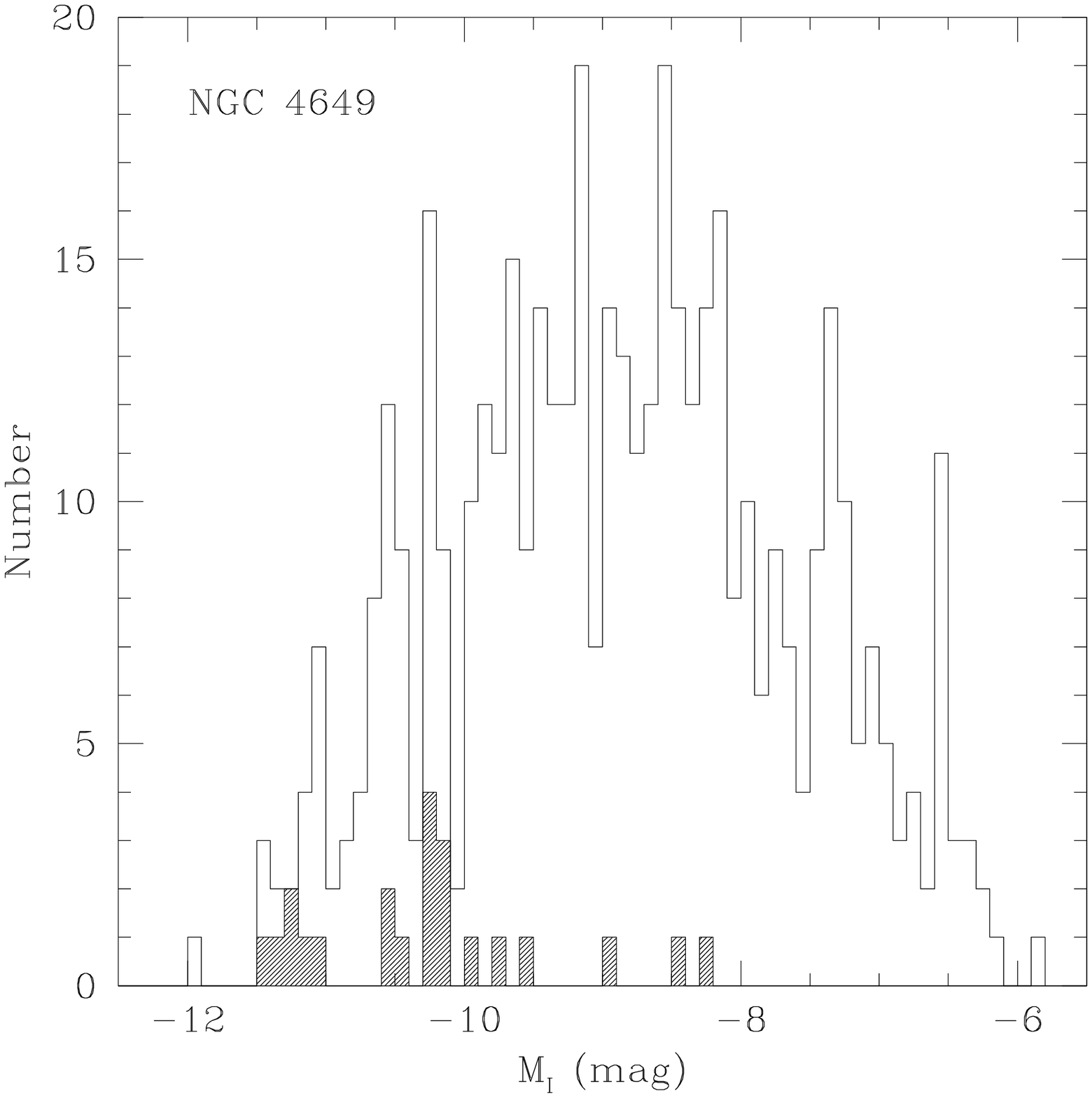}
\caption{Histograms of the number of globular clusters versus their
absolute magnitude for the galaxies NGC~4365 (left) and NGC~4649 (right).
The notation is the same as Figure~\protect\ref{fig:gc_mag}.
\label{fig:gc_mag2}}
\end{figure}

Figure~\ref{fig:gc_mag} shows histograms of the absolute I magnitude,
$M_I$, of the total GC sample (upper histogram) and of the GCs containing
LMXBs (shaded histogram) for the $L_X$ complete sample (left) and the
full source sample (right).
The results for the individual galaxies NGC~4365 and NGC~4649 are shown in
Figure~\ref{fig:gc_mag2}.
In NGC~1553 and NGC~4697, there are too few LMXBs associated with GCs to
allow a useful comparison.
All of the magnitudes were corrected for Galactic extinction.

The LMXBs seem to be associated preferentially with the more optically luminous
GCs.
For example, the median values of $M_I$ for non-X-ray GCs are
$-8.7$ ($L_X$ complete sample),
$-8.7$ (full sample),
$-9.1$ (NGC~4365),
and
$-8.8$ (NGC~4649),
while
the corresponding values for the X-ray GCs are
$-10.2$ ($L_X$ complete sample),
$-10.2$ (full sample),
$-10.1$ (NGC~4365),
and
$-10.3$ (NGC~4649).
Using the Wilcoxon or equivalent Mann-Whitney rank-sum tests
\citep{mw47},
the distribution of X-ray and non-X-ray luminosities are found to disagree
at more than the
5$\sigma$ ($L_X$ complete sample),
6$\sigma$ (full sample),
4$\sigma$ (NGC~4365),
and
5$\sigma$ (NGC~4649),
with the probability that they are drawn from the same distribution being
$<$$10^{-9}$ ($L_X$ complete sample),
$<$$10^{-11}$ (full sample),
$<$$10^{-4}$ (NGC~4365),
and
$<$$10^{-8}$ (NGC~4649).
We also compared the distributions of X-ray and non-X-ray GC absolute
magnitudes using the Kolmogorov-Smirnov (KS) two-sample test.
The probability that they were drawn from the same distribution is
less than
$10^{-8}$ ($L_X$ complete sample),
$10^{-9}$ (full sample),
$10^{-3}$ (NGC~4365),
$10^{-6}$ (NGC~4649).
Thus, the X-ray sources are strongly correlated with the brighter globular
clusters.
A similar result was found previously for the galaxy NGC~1399
by \citet{alm01} and for NGC~4472 by
\citet{kmz02}.

\begin{figure}[p]
\vskip4.0truein
\includegraphics{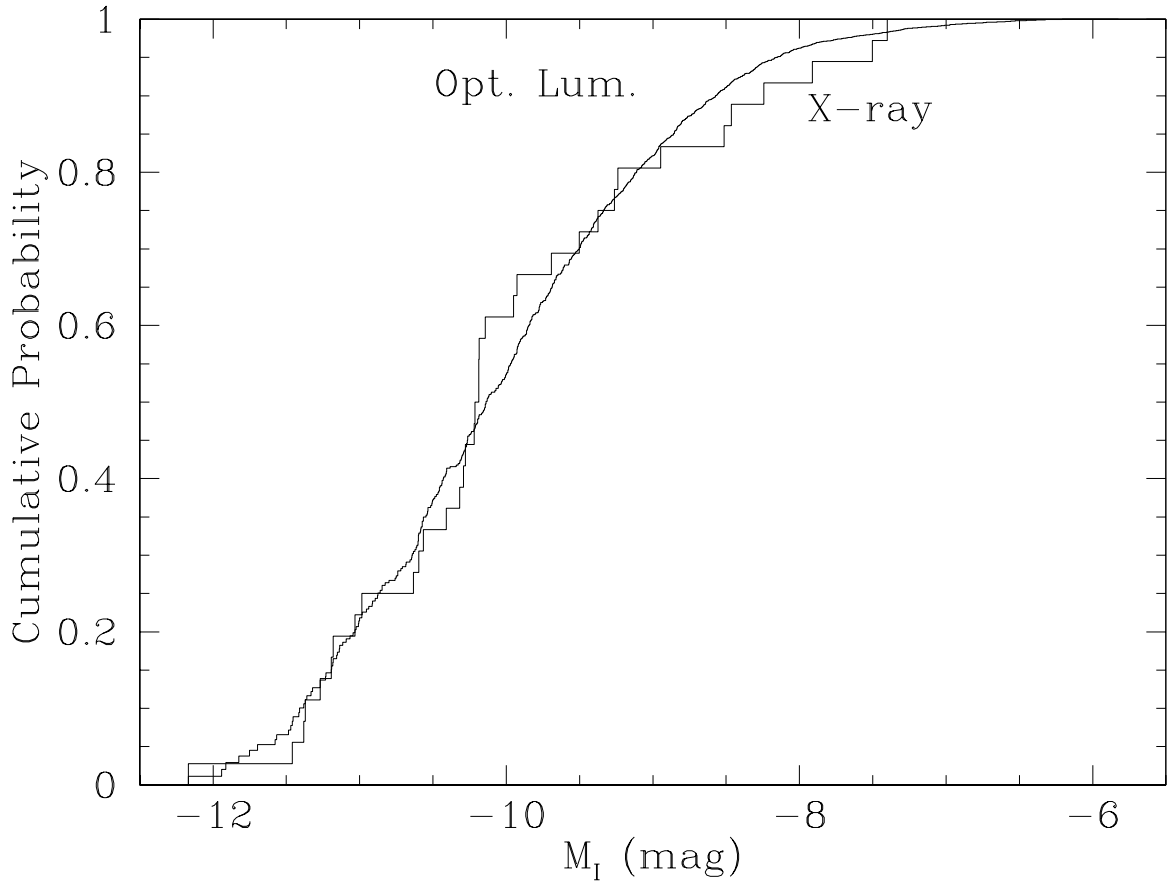}
\caption{Cumulative distribution functions for the probability that
a GC contains an X-ray source (``X-ray'') in the $L_X$ complete sample
and for the optical luminosity of GCs (``Opt.\ Lum.").
\label{fig:cpd_lum}}
\end{figure}

Of course, a correlation between optical luminosity and the probability
of having an X-ray source is not unexpected.
LMXBs contain normal stars, and globular clusters which have higher
luminosities have more stars as potential donors in LMXBs.
Thus, it is interesting to test the hypothesis that the probability that
a GC contains a LMXB is proportional to its optical luminosity.
Figure~\ref{fig:cpd_lum} compares the cumulative probability distribution of
LMXBs versus the cumulative distribution of the optical luminosity in
GCs (for the $L_X$ complete sample).
Both were accumulated starting at small $M_I$ (high optical luminosity) and
going to higher $M_I$ (low luminosity).
The optical luminosity is determined solely from the I-band absolute
magnitude.
The two cumulative distribution functions track one another fairly well.
For example,
half of the optical luminosity comes from GCs brighter than $M_I = -10.1$,
while the medium absolute magnitude of GCs with LMXBs is $-10.2$.
There is a slight tendency for the fainter GCs to have a higher probability
to have an X-ray source per unit luminosity, but it is not statistically
significant.
However, the GC samples are incomplete at the faint end, and the
completeness limit differs significantly from galaxy to galaxy in our
sample.
Although there is no obvious reason why this incompleteness would affect
GCs with X-ray sources differently than those without LMXBs,
this suggests caution in interpreting differences among faint GCs.
We used a modified version of the KS two-sample test
to compare the two distributions.
The largest vertical difference in the cumulative distributions
is $D = 0.115$, and this difference is not significant given the number
of LMXBs.
Very similar results were found for the full sample (albeit more strongly)
and for NGC~4365 and NGC~4649 (but less strongly).

Thus, the current data indicate that optically bright GCs are much more
likely to contain LMXBs than faint GCs, but the distribution is consistent
with a constant probability per unit optical luminosity.
A similar result was found previously for the galaxy NGC~4472 by \citet{kmz02}.
As noted in \S~\ref{sec:statistics}, the probability of having a LMXB per
unit optical luminosity of the GC also appears to be constant from galaxy
to galaxy, with a value of about
$1.5 \times 10^{-7}$ LMXBs per $L_{\odot,I}$ for
$L_X \ga 1 \times 10^{38}$ erg s$^{-1}$ (0.3--10 keV).
The value rises to about
$2.0 \times 10^{-7}$ LMXBs per $L_{\odot,I}$ at lower X-ray luminosities
($L_X \ga 3 \times 10^{37}$ erg s$^{-1}$).

\subsection{Optical Colors} \label{sec:gc_color}

\begin{figure}
\plottwo{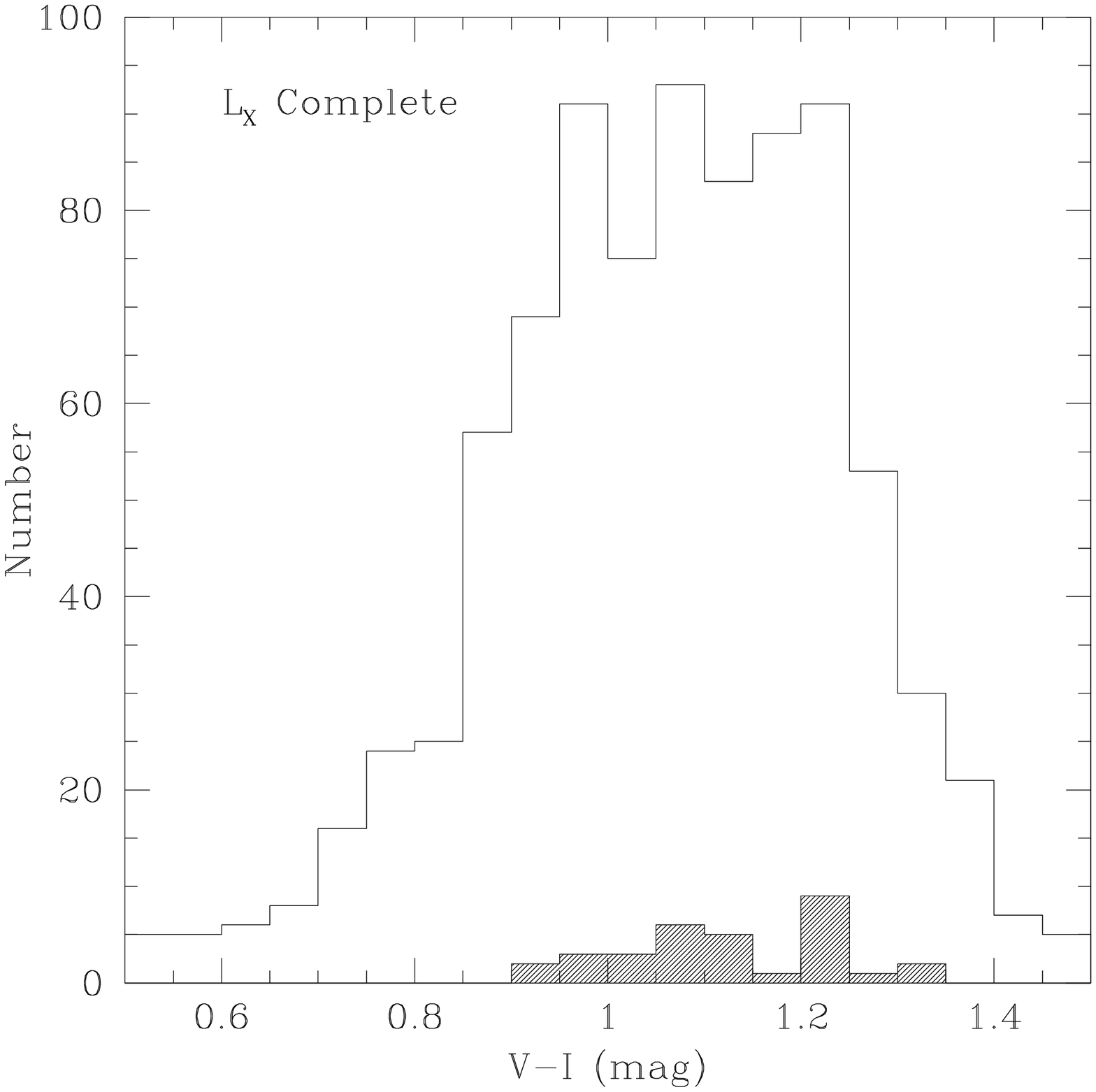}{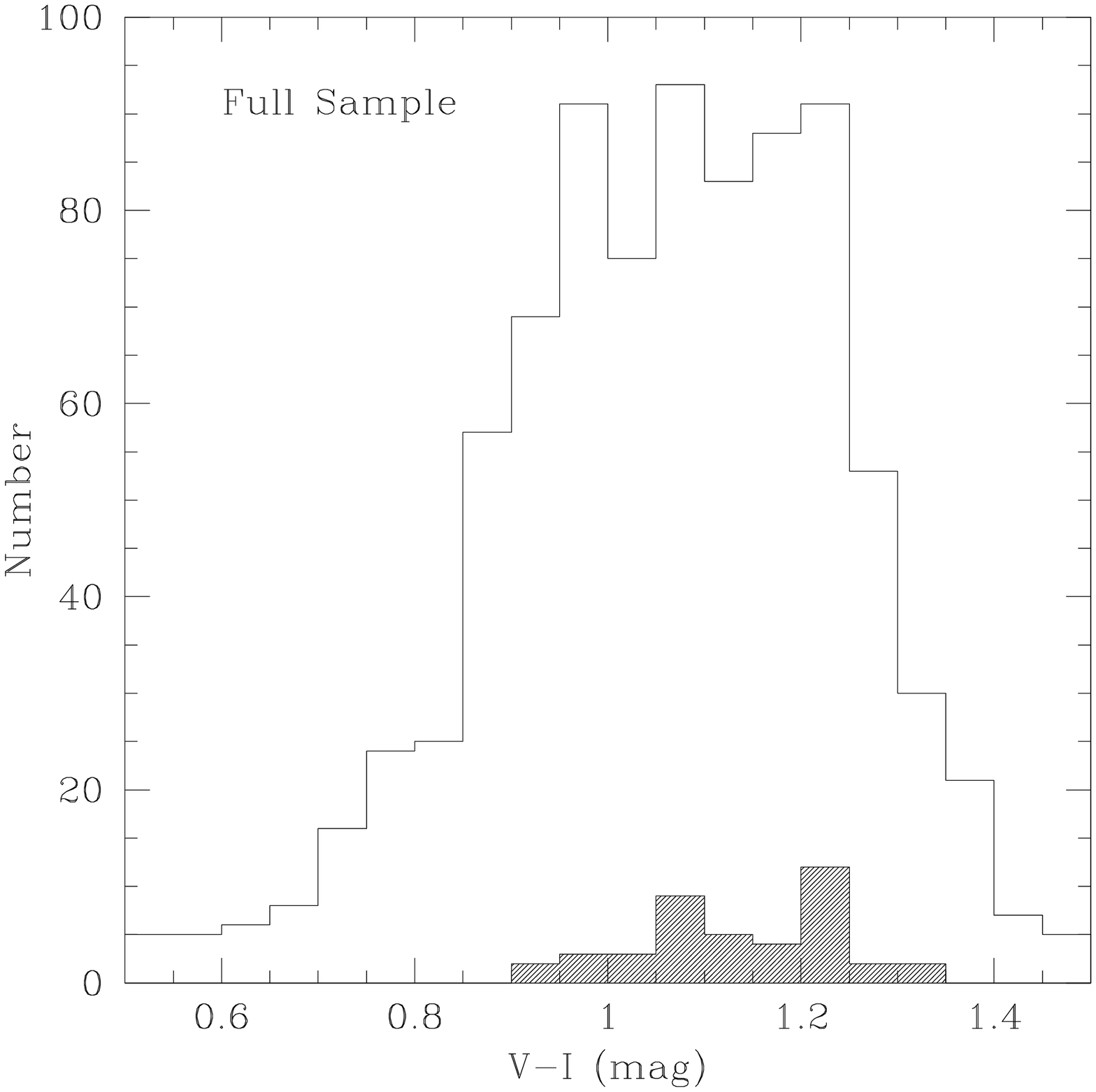}
\caption{Histograms of the number of globular clusters versus their
optical color, $V-I$.
The left panel is for the $L_X$ complete sample, while the right panel is
for the full sample.
In each case, 
the upper histogram is for all of the GCs in the galaxies.
The lower shaded histogram shows the GCs which contain identified LMXBs.
The histogram bins are 0.1 mag wide.
\label{fig:gc_color}}
\end{figure}

\begin{figure}
\plottwo{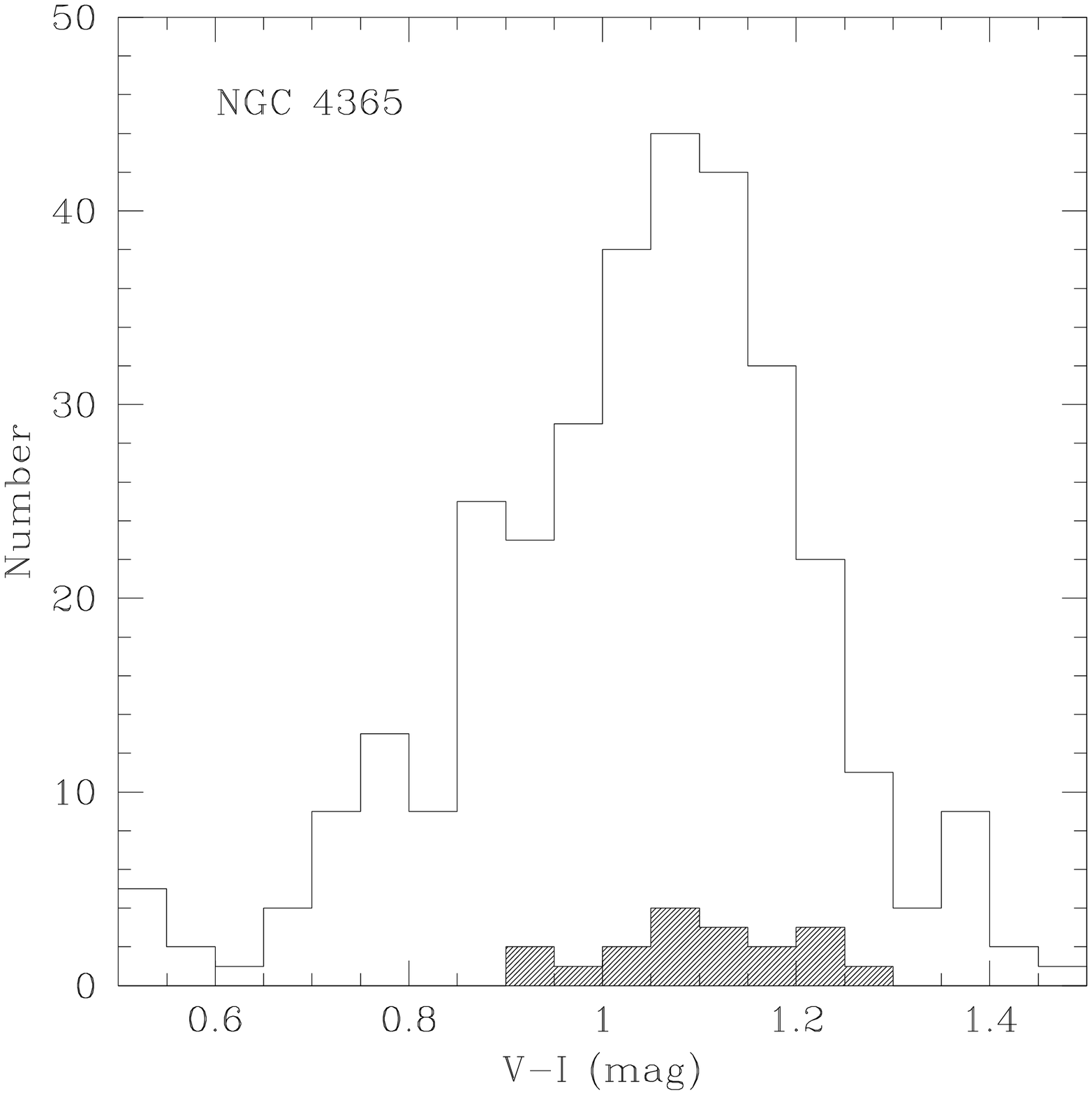}{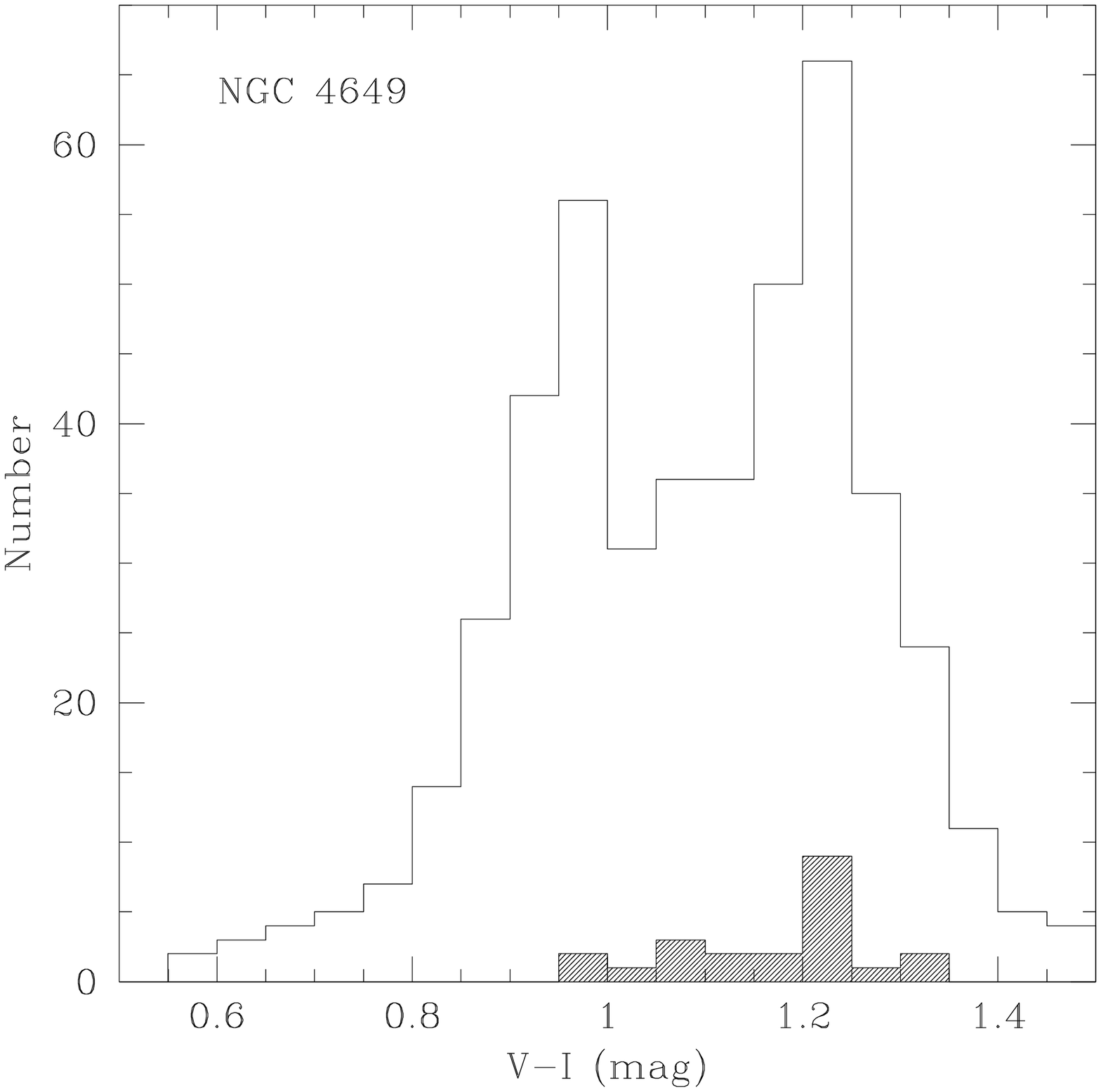}
\caption{Histograms of the number of globular clusters versus their
optical color for the galaxies NGC~4365 (left) and NGC~4649 (right).
The notation is the same as Figure~\protect\ref{fig:gc_color}.
\label{fig:gc_color2}}
\end{figure}

Figure~\ref{fig:gc_color} shows histograms of the $V-I$ colors
for the total GC sample (upper histogram) and for the GCs containing
LMXBs (shaded histogram) for the $L_X$ complete sample (left) and the
full source sample (right).
The colors are all corrected for Galactic reddening.
Because these samples contain a mixture of three galaxies,
the overall color distribution may be less obviously bimodal than that seen
in some individual galaxies
\citep[e.g.,][]{kmz02}.
The results for the individual galaxies NGC~4365 and NGC~4649 are shown in
Figure~\ref{fig:gc_color2}.
As has been noted previously \citep[e.g.,][]{kmz02}, NGC~4649 has a
bimodal color distribution, whereas NGC~4365 was a broad color distribution
which is not clearly bimodal.
In NGC~1553, there are too few LMXBs associated with GCs to allow a useful
comparison, and no colors were available to us for the GCs in NGC~4697.

The LMXBs appear to be associated preferentially with the redder
GCs (larger values of $V-I$), although the trend is not as strong as
that for the absolute magnitude or optical luminosity (\S~\ref{sec:gc_mag}).
The median colors of the non-X-ray GCs are
$V-I = 1.07$ ($L_X$ complete sample),
$V-I = 1.07$ (full sample),
$V-I = 1.05$ (NGC~4365),
and
$V-I = 1.09$ (NGC~4649),
while the corresponding medians for the X-ray GCs are
$1.12$ ($L_X$ complete sample),
$1.14$ (full sample),
$V-I = 1.09$ (NGC~4365),
and
$V-I = 1.21$ (NGC~4649).
Using the Wilcoxon or Mann-Whitney rank-sum tests,
the probabilities that the distribution of X-ray and non-X-ray optical
colors were drawn from the same distribution are
$0.018$ ($L_X$ complete sample),
$0.0019$ (full sample),
$0.046$ (NGC~4365),
and
$0.0069$ (NGC~4649),
which correspond to differences of
2.1$\sigma$ ($L_X$ complete sample),
2.9$\sigma$ (full sample),
1.7$\sigma$ (NGC~4365),
and
2.5$\sigma$ (NGC~4649).
Using the KS test, the probabilities that the two color distributions
where drawn from the same distribution are
$0.030$ ($L_X$ complete sample),
$0.0013$ (full sample),
$0.14$ (NGC~4365),
and
$0.031$ (NGC~4649).
Thus, a significant trend for LMXBs to be associated with red GCs appears
in the $L_X$ complete sample, in the full sample, and in NGC~4649.
A similar correlation with the optical color of GCs was found previously
for the galaxy NGC~1399 by \citet{alm01}
and for NGC~4472 by \citet{kmz02}.
The trend is less obvious in NGC~4365.
Recent optical/IR observations of GCs in NGC~4365 indicate that this galaxy
has both old and intermediate age ($\sim$5 Gyr) red GCs
\citep{PZK+02,LBB+03},
which may account for its broad but unimodal color distribution.
The LMXBs seem to be preferentially associated with the younger red GCs
\citep{SAS03}.

\begin{figure}
\epsscale{1.0}
\plotone{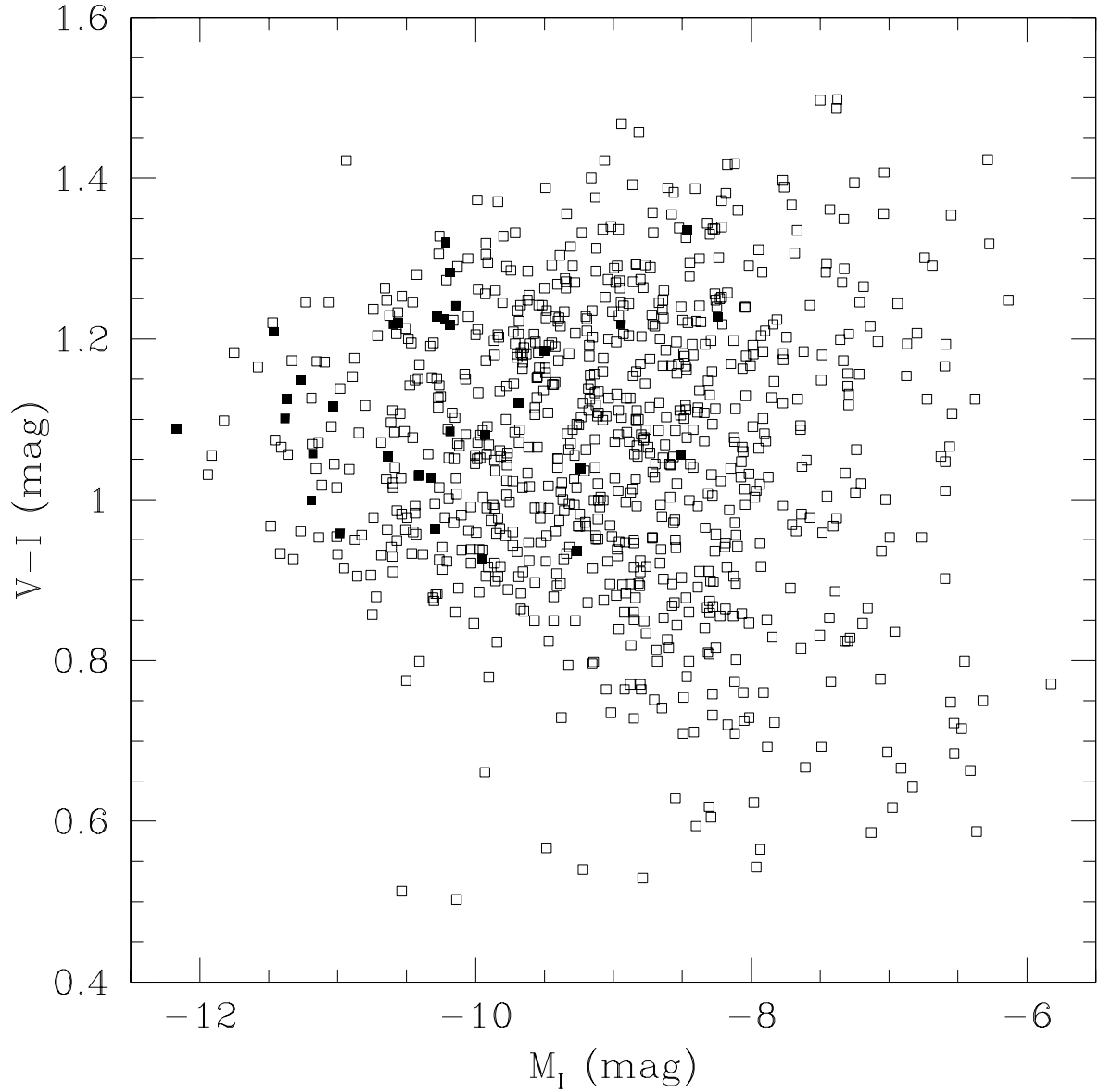}
\caption{Scatter plot of the absolute magnitudes $M_I$ and 
optical colors $V-I$ for the GCs.
The open squares are non-X-ray GCs, while the filled squares are
X-ray GCs in the $L_X$ complete sample.
\label{fig:gc_color_scatter}}
\end{figure}

A correlation between LMXBs and GC color might result from the correlation
between LMXBs and
the optical luminosity of GCs, if the brighter globular clusters were
preferentially redder.
Figure~\ref{fig:gc_color_scatter} shows the scatter diagram between the optical
colors and absolute magnitudes for the GCs, with those associated with
LMXBs in the $L_X$ complete sample indicated.
Various correlation measures show no evidence for a significant statistical
correlation between absolute magnitude and color.
This test doesn't take into account the various selection effects
which went into constructing this sample;
it only shows that there is no strong correlation between absolute
magnitude and color within this sample.
However, previous detailed optical studies, including selection
effects, on the GCs in these and other
early-type galaxies have found no significant correlation of color with
optical luminosity
\citep[e.g.,][]{kw01a}.

The correlations between X-ray sources and GC color would
seem to suggest that the evolution of X-ray binaries in GCs is
affected either by the metallicity or age of the GC, with younger and/or
more metal-rich GCs having more LMXBs.

\section{X-ray Properties of GCs} \label{sec:xray}

\subsection{X-ray Luminosities} \label{sec:xray_lum}

\begin{figure}
\epsscale{1.0}
\plotone{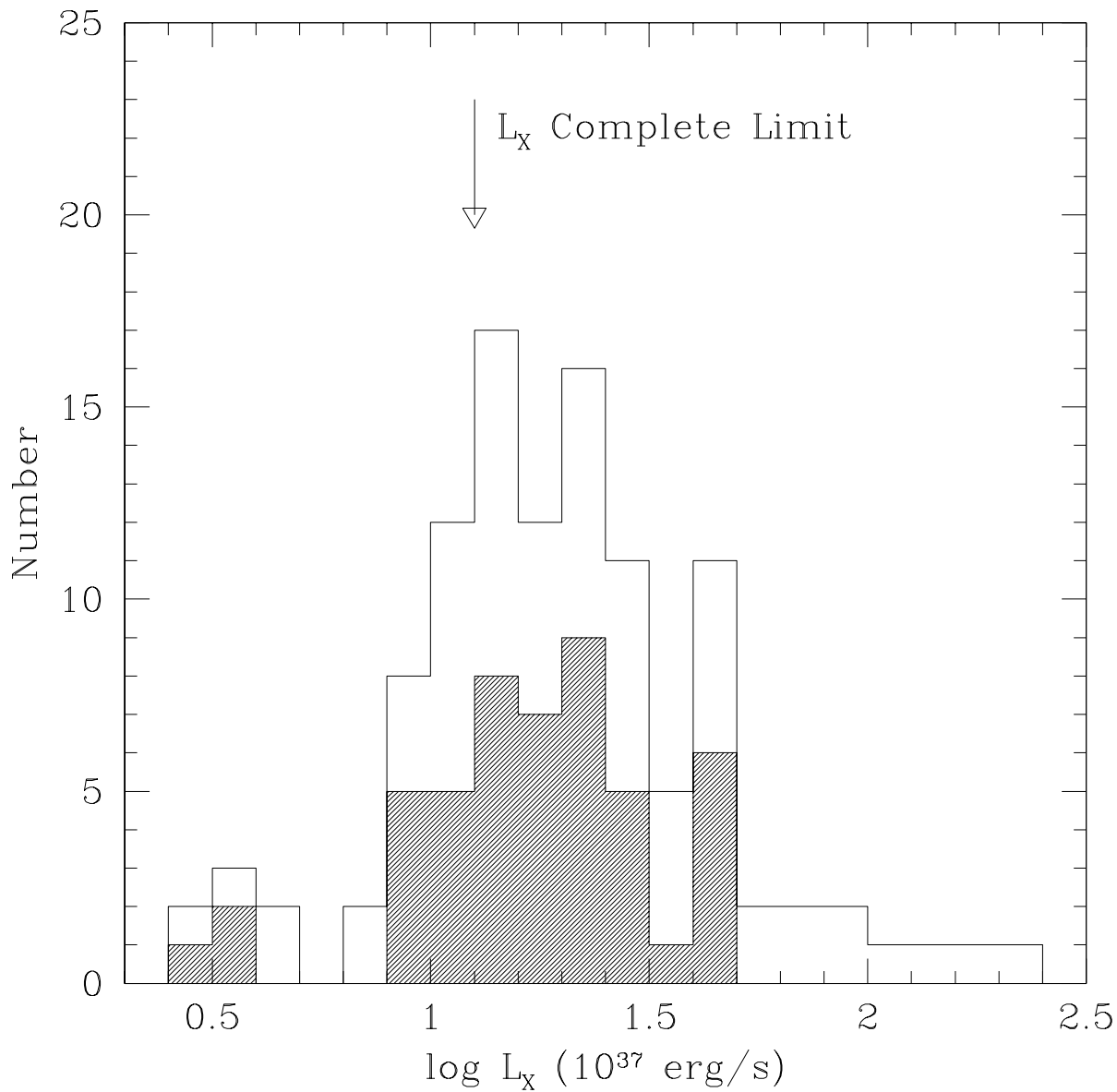}
\caption{Histograms of the number of X-ray sources versus their
X-ray luminosity (unabsorbed 0.3--10 keV).
The upper histogram is for all of the sources in our full sample.
The lower shaded histogram shows the X-ray sources identified with GCs.
The histogram bins are 0.1 dex wide.
The arrow marks the limiting X-ray luminosity above which the sample is
complete.
\label{fig:x_lum}}
\end{figure}

\begin{figure}
\plottwo{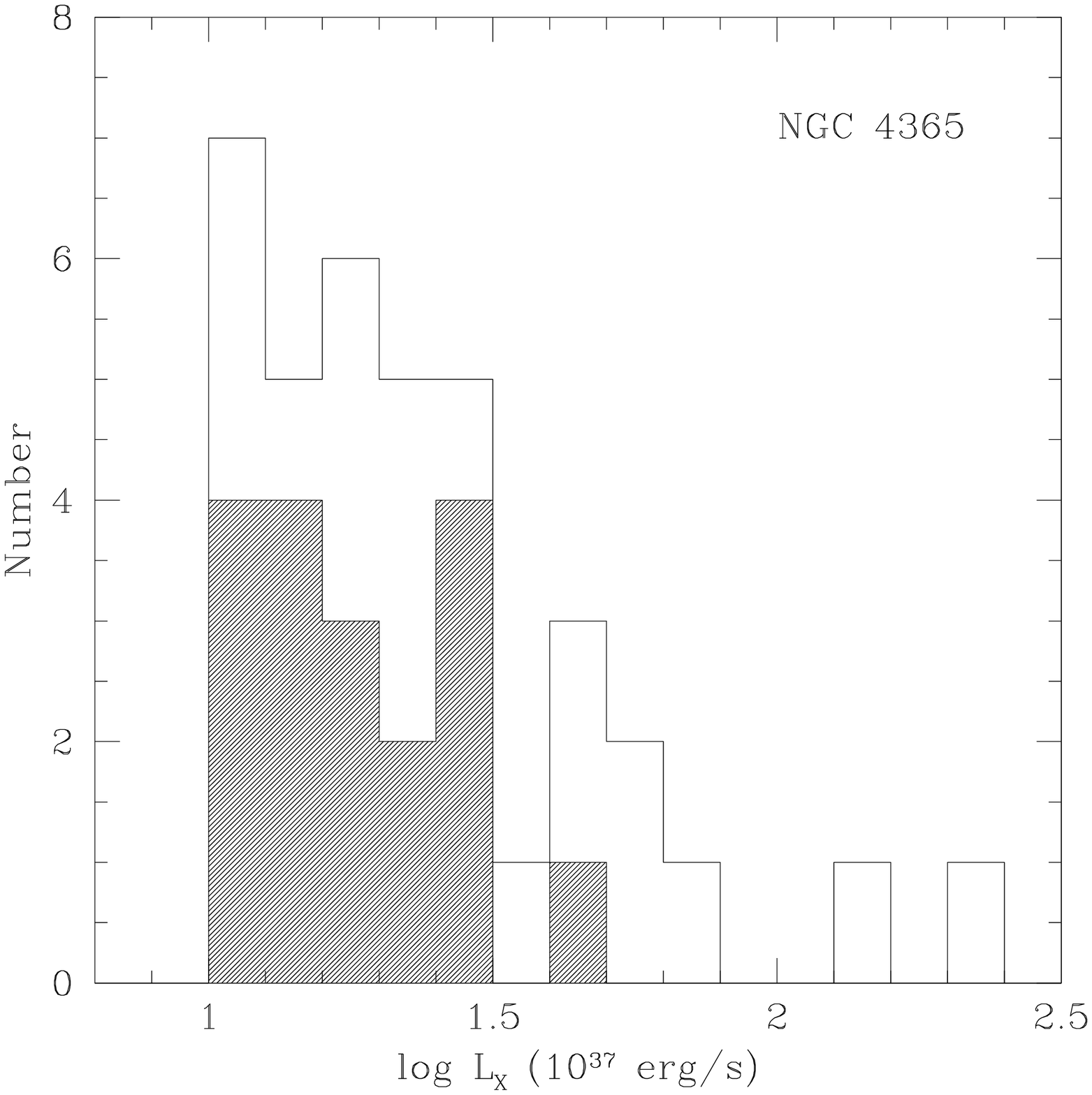}{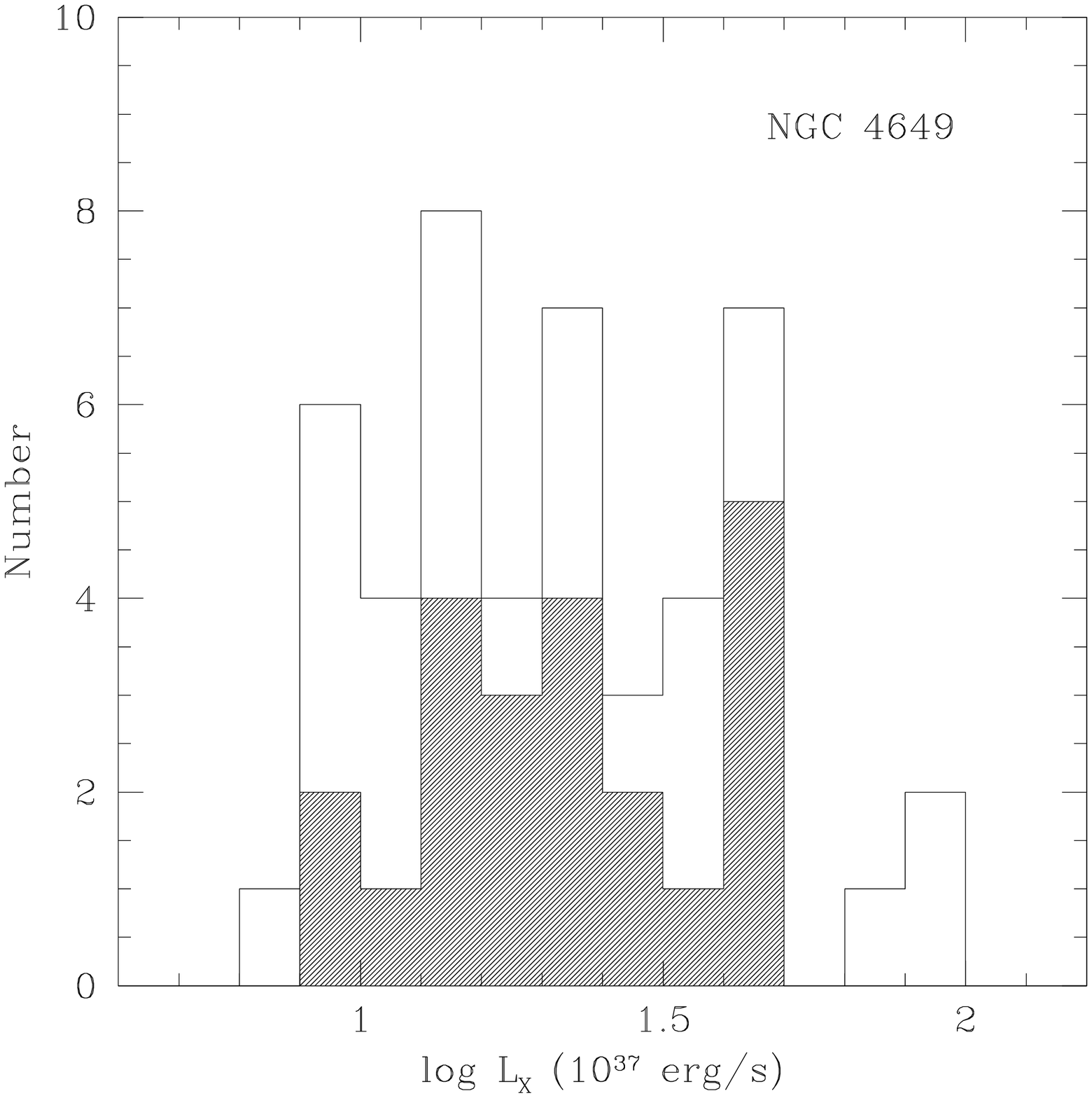}
\caption{Histograms of the number of X-ray sources versus their
X-ray luminosity (unabsorbed 0.3--10 keV)
for the galaxies NGC~4365 (left) and NGC~4649 (right).
The notation is the same as Figure~\protect\ref{fig:x_lum}.
\label{fig:x_lum2}}
\end{figure}

Figure~\ref{fig:x_lum} shows histograms of the numbers of X-ray sources
as a function of their X-ray luminosities for all of the sources
(upper histogram) and for the sources associated with GCs.
This figure is for the full sample, but the lower limit luminosity for
the $L_X$ complete sample is also marked.
Figure~\ref{fig:x_lum2} shows the distribution of X-ray luminosities for
the individual galaxies NGC~4365 and NGC~4649;
NGC~1553 and NGC~4697 had too few GC-LMXBs to provide useful comparisons.
The X-ray luminosities are in the band 0.3--10 keV and have been corrected
for absorption.
Although the statistical accuracy is limited by the small number X-ray sources,
at most luminosity levels there is no obvious difference in the two
distributions.
However, there may be a tendency for the most X-ray-luminous
sources to avoid GCs.
In the $L_X$ complete sample.
the rank sum test and KS test both find that the GC sources are fainter.
The probability that the two populations are drawn from the
same distribution can be rejected at the 97\% level, corresponding
to a difference of about 1.9$\sigma$.

It is generally found that the luminosity functions of X-ray sources
in early-type galaxies have a broken power-law form, with a break
luminosity of $L_b \sim 3 \times 10^{38}$ erg s$^{-1}$
\citep[e.g.,][]{sib01}.
This break luminosity is close to the Eddington luminosity of
a 1.4 $M_\odot$ neutron star.
As a result, \citet{sib00,sib01} have argued that the sources with
$L_X > L_b$ predominantly contain black holes, while those below this
luminosity may contain neutron stars.
In the $L_X$ complete sample.
19$\pm$6\% of LMXBs in GC have luminosities greater than
$3 \times 10^{38}$ erg s$^{-1}$, while the fraction for the non-GC
sources is 44$\pm$8\% (1$\sigma$ error bars).
Thus, the difference is somewhat significant.

Our results are similar to those found for the X-ray bright elliptical
galaxy NGC~4472 by \citet{kmz02}, who found no significant difference in
the X-ray luminosities of GC and non-GC sources.
On the other hand,
\citet{alm01} found that the brightest X-ray sources in NGC~1399
were mainly associated with GCs, which is the opposite of the
weak tendency we find.

\subsection{X-ray Hardness Ratios} \label{sec:xray_colors}

We used hardness ratios or X-ray ``colors'' to crudely characterize
the spectra of the sources.
We defined two hardness ratios as
\begin{eqnarray}
H21 & \equiv & \frac{M - S}{M + S} \, , \label{eq:h21} \\
H31 & \equiv & \frac{H - S}{H + S} \, , \label{eq:h31}
\end{eqnarray}
where $S$, $M$, and $H$ are the total counts in the
soft (0.3--1 keV),
medium (1--2 keV),
and
hard (2--10 keV)
bands, respectively.
Although all of the galaxies lie in directions with low Galactic
absorbing columns $N_H$ (Table~\ref{tab:galaxies}),
there is a small variation in the absorption towards the different
galaxies.
Thus, we have corrected the hardness ratio to remove the effects
of Galactic absorption.
Let $S_0$, $M_0$, and $H_0$ be the soft, medium, and hard band counts,
corrected for absorption.
We define
$f21 \equiv ( S / S_0 ) / ( M / M_0 )$, and
$f31 \equiv ( S / S_0 ) / ( H / H_0 )$.
Then, the absorption-corrected hardness ratios are
\begin{eqnarray}
H21^0 & = &
\frac{ ( 1 + H21 ) f21 - ( 1 - H21 )}{ ( 1 + H21 ) f21 + ( 1 - H21 )}
\, , \label{eq:h21corr} \\
H31^0 & = &
\frac{ ( 1 + H31 ) f31 - ( 1 - H31 )}{ ( 1 + H31 ) f31 + ( 1 - H31 )}
\, . \label{eq:h31corr}
\end{eqnarray}
We calculated the corrections assuming a single spectrum for all of the
sources, which we took to be 7 keV thermal bremsstrahlung.
The corrections are smaller than the errors in the hardness ratios for
almost all of the sources, and the differences in the corrections
from galaxy to galaxy are even smaller. 

\begin{figure}
\epsscale{1.0}
\plotone{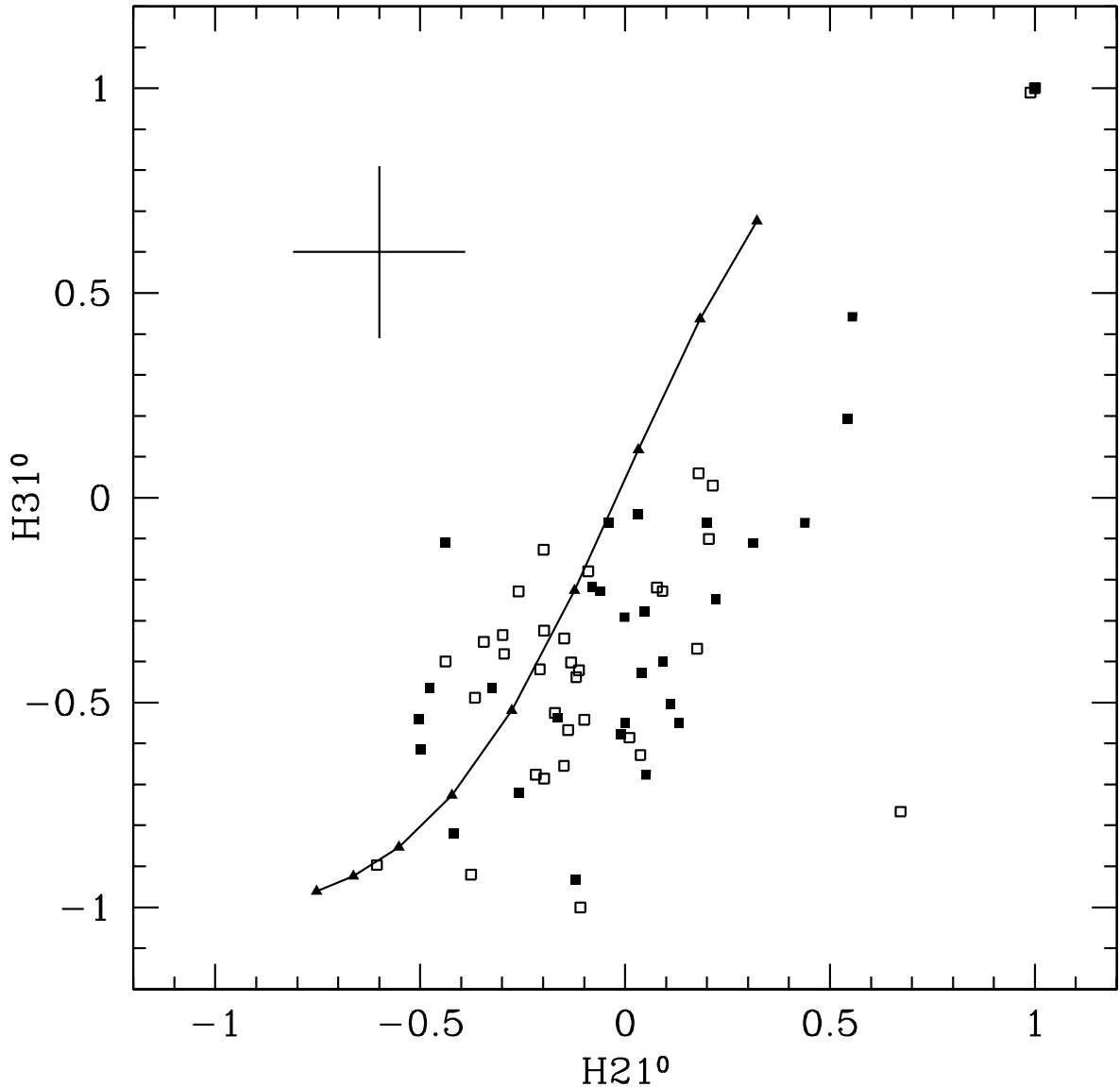}
\caption{Plot of the Galactic absorption-corrected X-ray hardness ratios,
$H21^0$ and $H31^0$, for the X-ray sources with at least 20 net counts
in the $L_X$ complete sample.
The open squares are non-GC sources, while the filled squares are
sources identified with GCs.
The solid line and triangles show the hardness ratios for power-law
spectral models, where the triangles indicate values of the power-law
photon number index
of $\Gamma = 0$ (upper right) to 3.2 (lower left) in increments of
0.4.
The error bars at the upper left give the approximate uncertainties for
an average source (one with about 50 net counts).
\label{fig:x_color}}
\end{figure}

\begin{figure}
\plottwo{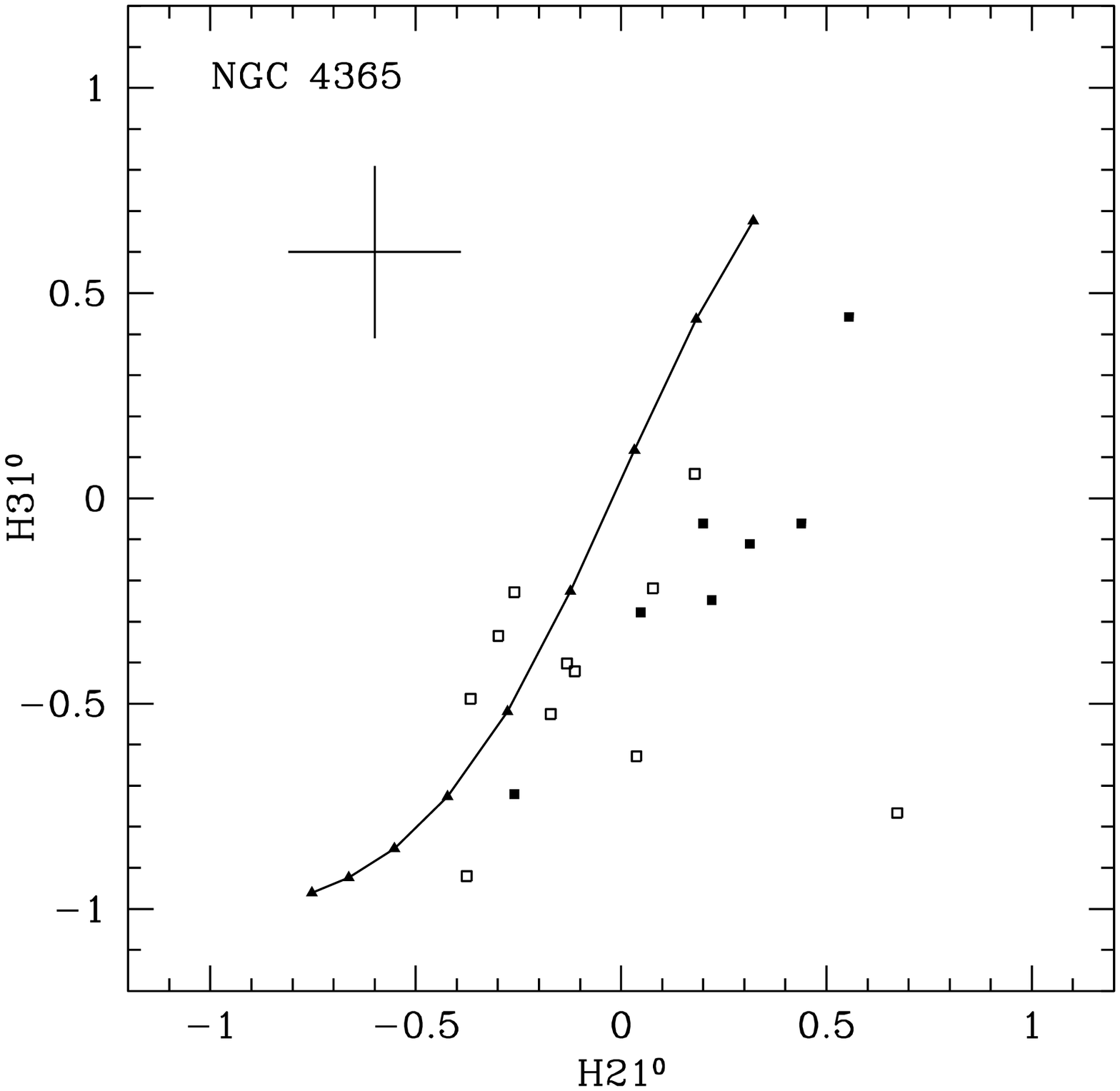}{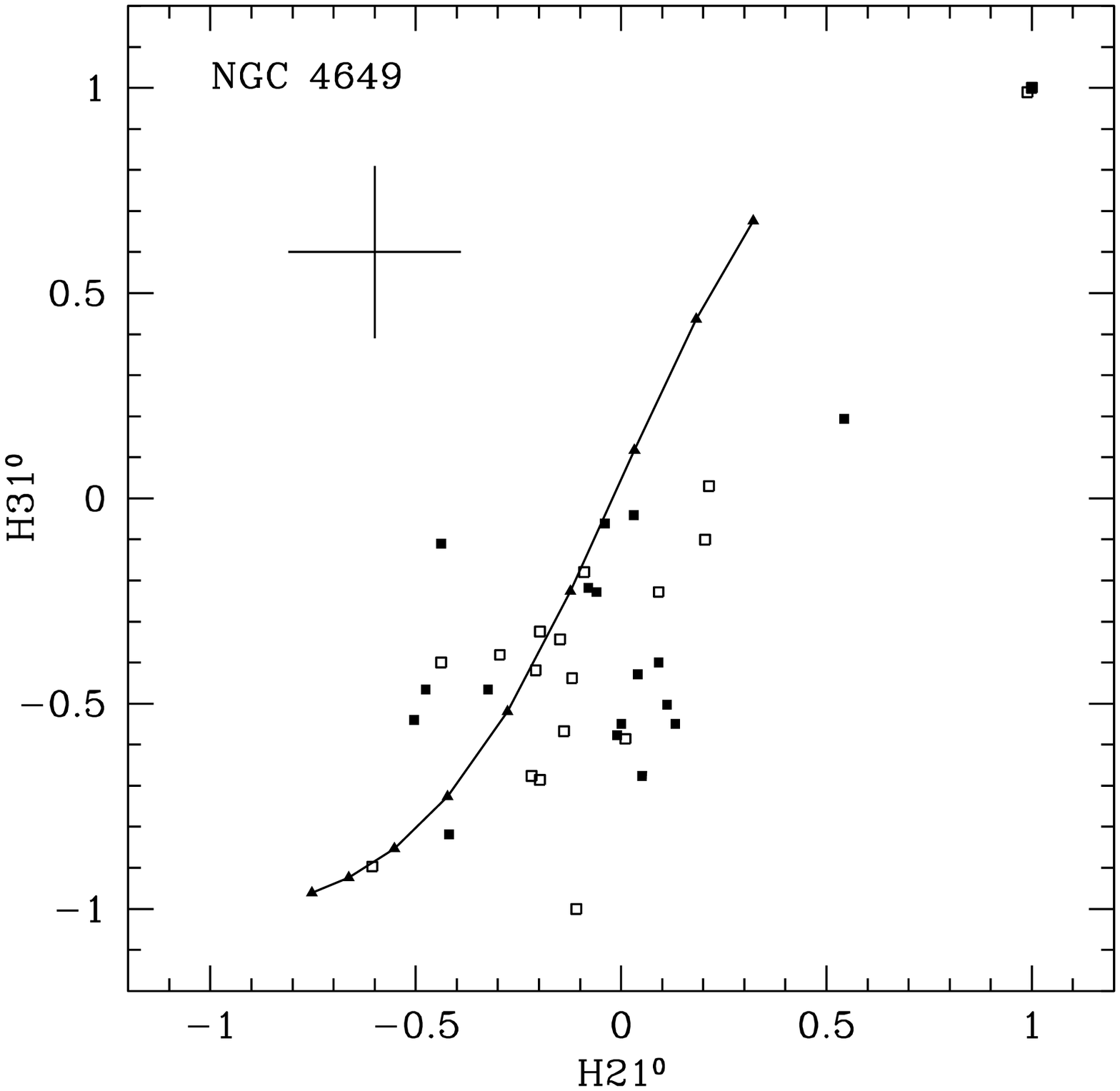}
\caption{Plots of the X-ray hardness ratios $H21^0$ and $H31^0$
for the X-ray sources 
in the galaxies NGC~4365 (left) and NGC~4649 (right).
The notation is the same as Figure~\protect\ref{fig:x_color}.
\label{fig:x_color2}}
\end{figure}

In Figure~\ref{fig:x_color}, we show the absorption-corrected hardness
ratios for the sources with more than 20 net counts in the $L_X$ complete
sample.
The filled squares are the sources identified with GCs, while
the open squares are non-GC sources.
For comparison,
the solid line shows the hardness ratios for power-law
spectral models;
the triangles indicate values of the power-law photon number index
of $\Gamma = 0$ (upper right) to 3.2 (lower left) in increments of
0.4.
Although there are significant variations in spectra from source to
source, on average the source spectrum can be fit by either thermal
bremsstrahlung with a temperature of $\sim$7 keV or a power-law with
a photon index $\Gamma \sim 1.5$ plus Galactic absorption.
The hardnesses ratios for the sources in the galaxies NGC~4365 and NGC~4649
are given separately in Figure~\ref{fig:x_color2}.
NGC~1553 and NGC~4697 had too few GC-LMXBs to provide useful comparisons.

\begin{figure}
\vskip4.0truein
\includegraphics{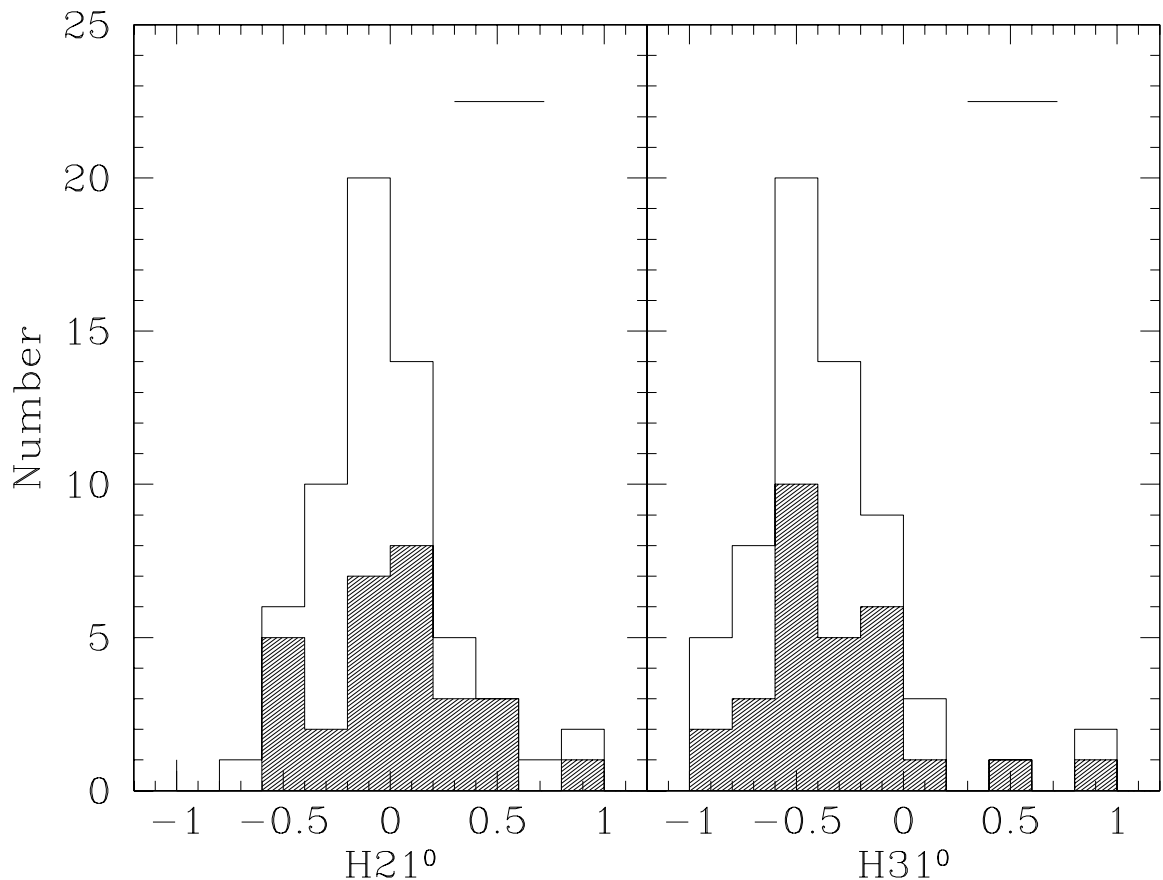}
\caption{Histograms of the number of X-ray sources versus their
Galactic absorption-corrected X-ray hardness ratios
$H21^0$ (right panel)
and $H31^0$ (left panel).
Only sources with at least 20 net counts in the $L_X$ complete sample
are included.
In both panels,
the upper histogram is for all of the sources.
The lower shaded histogram shows the X-ray sources identified with GCs.
The histogram bins are 0.2 wide.
The horizontal bar at the top of each figures shows the typical uncertainty
for that hardness ratio (the error bar for a source with about 50 net
counts).
\label{fig:x_hard}}
\end{figure}

Histograms of the $H21^0$ and $H31^0$ hardness ratios are given in
Figure~\ref{fig:x_hard} for the $L_X$ complete sample.
Considering both Figures~\ref{fig:x_color} and \ref{fig:x_hard},
no very significant differences in the hardness ratios are evident.
The rank-sum and KS tests suggest that the $H21^0$ colors of the
GC sources are slightly harder (larger $H21^0$), but the difference
is only significant at the 93\% confidence level.
The same tests indicate that the $H31^0$ distributions are identical.
Thus, there is no strong evidence that the spectra of the GC and
non-GC sources differ in any systematic way.
This agrees with the conclusions of \citet{alm01} and \citet{kmz02} for
NGC~1399 and NGC~4472, respectively.

There are two X-ray sources (one GC and one non-GC source) with hardness
ratios of \linebreak
$(H21^0,H31^0) \approx (1,1)$.
It is likely that these X-ray sources are strongly self-absorbed.
In some cases, such sources are found to be identified with background AGNs
\citep[e.g.,][]{sib00}.
However, both of these X-ray sources are projected within the central regions
of NGC~4649, where the chance of finding a background source is low
(due to the small solid angle involved).
One might be concerned that the GC optical identification of one of the hard
sources might actually be a background AGN.
However, the identified GC has very normal red optical colors, which would
be atypical (but not unheard of) for an AGN.
This suggests that a small portion of the GC and non-GC X-ray sources
may have strongly self-absorbed spectra.

In some early-type galaxies, ``supersoft'' sources have been found which
have no X-ray emission at energies above 1 keV
\citep[e.g.,][]{sib00}.
These sources can be fit with $\sim$75 eV blackbody spectra.
In terms of their spectra, these sources are similar to the many supersoft
sources observed in our Galaxy and in M31
\citep[e.g.,][]{kvdh97}.
In general, the Galactic supersoft sources are believed to be accreting
white dwarf stars in binaries, with the luminosity due to steady nuclear
burning.
However, the supersoft sources found in distant early-type galaxies have
very large bolometric luminosities which exceed the Eddington luminosity
for a Chandrasekhar mass white dwarf.
A similar bright supersoft source has been found in the bulge of M81
\citep{sgs+02}.
One hypothesis is that these sources contain intermediate mass
($\sim 10^2 - 10^3 \, M_\odot$) accreting black holes
\citep{sgs+02}.
It is unfortunate that there are no supersoft sources in the regions of
these galaxies which are covered by the GC surveys,
particularly as it has been suggested that intermediate mass black holes
are produced in GCs \citep{mh02}.

\subsection{X-ray Variability} \label{sec:xray_vary}

During these {\it Chandra} observations lasting typically $\sim$10 hours,
about 10\% of the X-ray sources are found to vary significantly.
Due to the weakness of the sources and some features of the detector
used, only secular variations over the observation period (source
turns on, turns off, or increases or decreases by a significant fraction)
are detectable.
Of the 111 LMXBs in our full sample, 9 varied significantly, of which
4 were in GCs.
In the $L_X$ complete sample, 4 sources varied, of which 2 were in GCs.
In the entire full sample, 44\% of the LMXBs are in GCs.
Within these very limited statistics, there is no evidence for a
strong difference in the variability of GC and non-GC X-ray sources.

\subsection{X-ray/Optical Correlations} \label{sec:xray_opt}

\begin{figure}
\epsscale{1.0}
\plotone{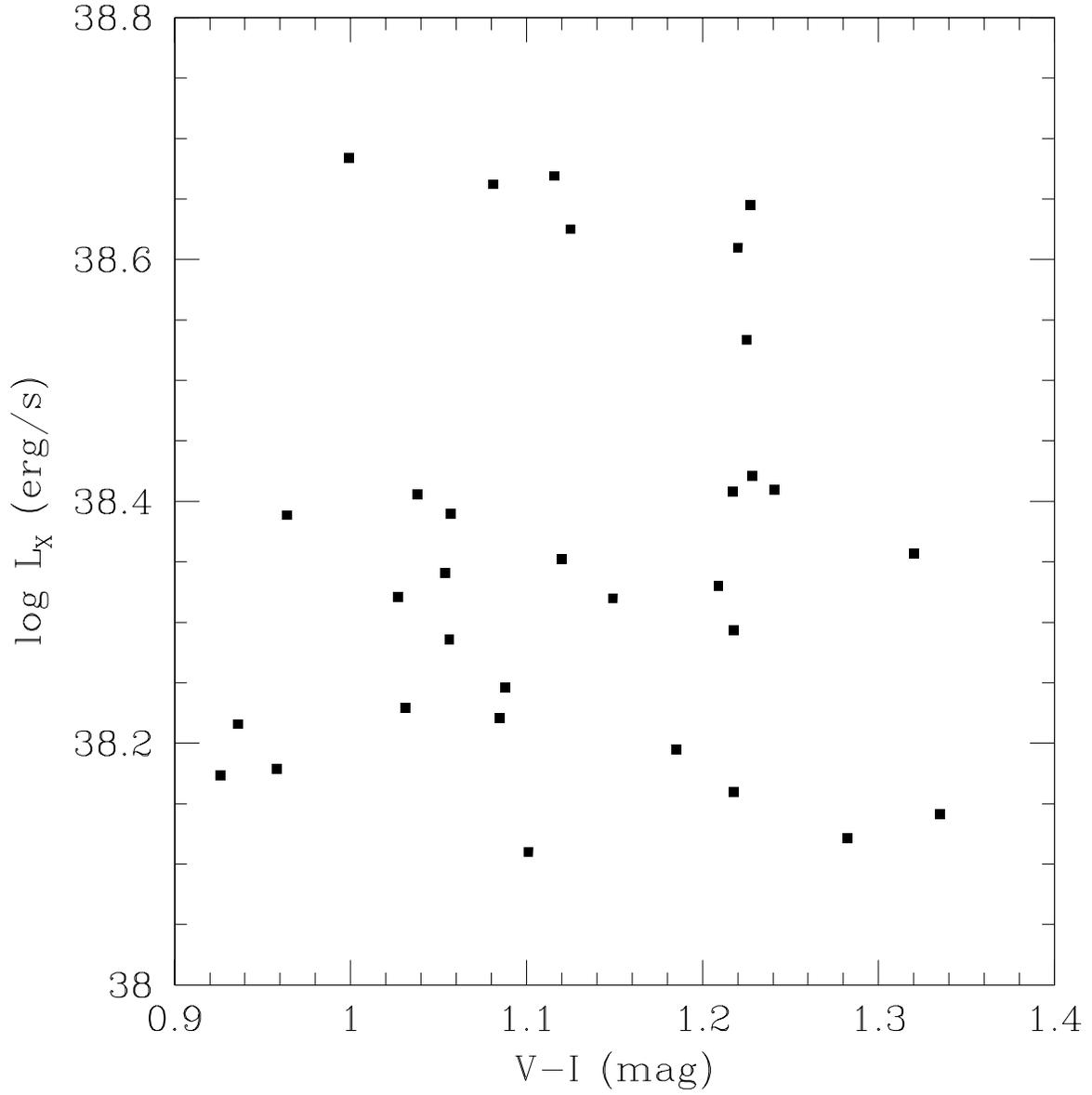}
\caption{X-ray luminosity of LMXBs located in GCs versus the optical color
of the GC.
All sources are in the $L_X$ complete sample.
No significant correlation is evident.
\label{fig:vi_lx}}
\end{figure}

Among the LMXBs which were located in GCs, we also searched for
correlations between their X-ray properties ($L_X$, $H21^0$, $H31^0$)
and the optical properties of the GCs in which they are situated.
For example, Figure~\ref{fig:vi_lx} shows the logarithm of the X-ray
luminosity ($\log L_X$) plotted against the optical color ($V - I$) of
the GC for sources in the $L_X$ complete sample.
No significant correlations were found between any of the X-ray
properties and any of the optical properties discussed here.
The probabilities that the properties were uncorrelated were
all 24\% or greater for the various pairs of properties.
One possible problem with these tests is the relatively small sample
of only 49 LMXBs in GCs.

\section{Spatial Distributions} \label{sec:spatial}

As discussed in \S~\ref{sec:intro}, the available GC samples
for these galaxies only cover a small fraction of the areas of the
galaxies.
As a result of this spatial incompleteness and possible position-dependent
selection effects for both the GC and LMXBs, we will not analyze the
spatial distribution functions of the GCs or LMXBs.
Instead, we will only address the relative distributions of GC and
non-GC LMXBs, and of X-ray and non-X-ray GCs.
Also, we will only consider the radial distribution of sources.
Because the galaxies have differing eccentricities, it would be difficult
to compare their azimuthal distributions.
In order to improve the statistics by combining the data from all of
the galaxies, we will scale the projected radius $r$ of each source
by the effective radius $r_{\rm eff}$ for its host galaxy.
The adopted values of $r_{\rm eff}$ are given in Table 1.
The references are given in the papers containing the X-ray data.

\subsection{Distribution of LMXBs} \label{sec:spatial_xray}

\begin{figure}
\epsscale{1.0}
\plotone{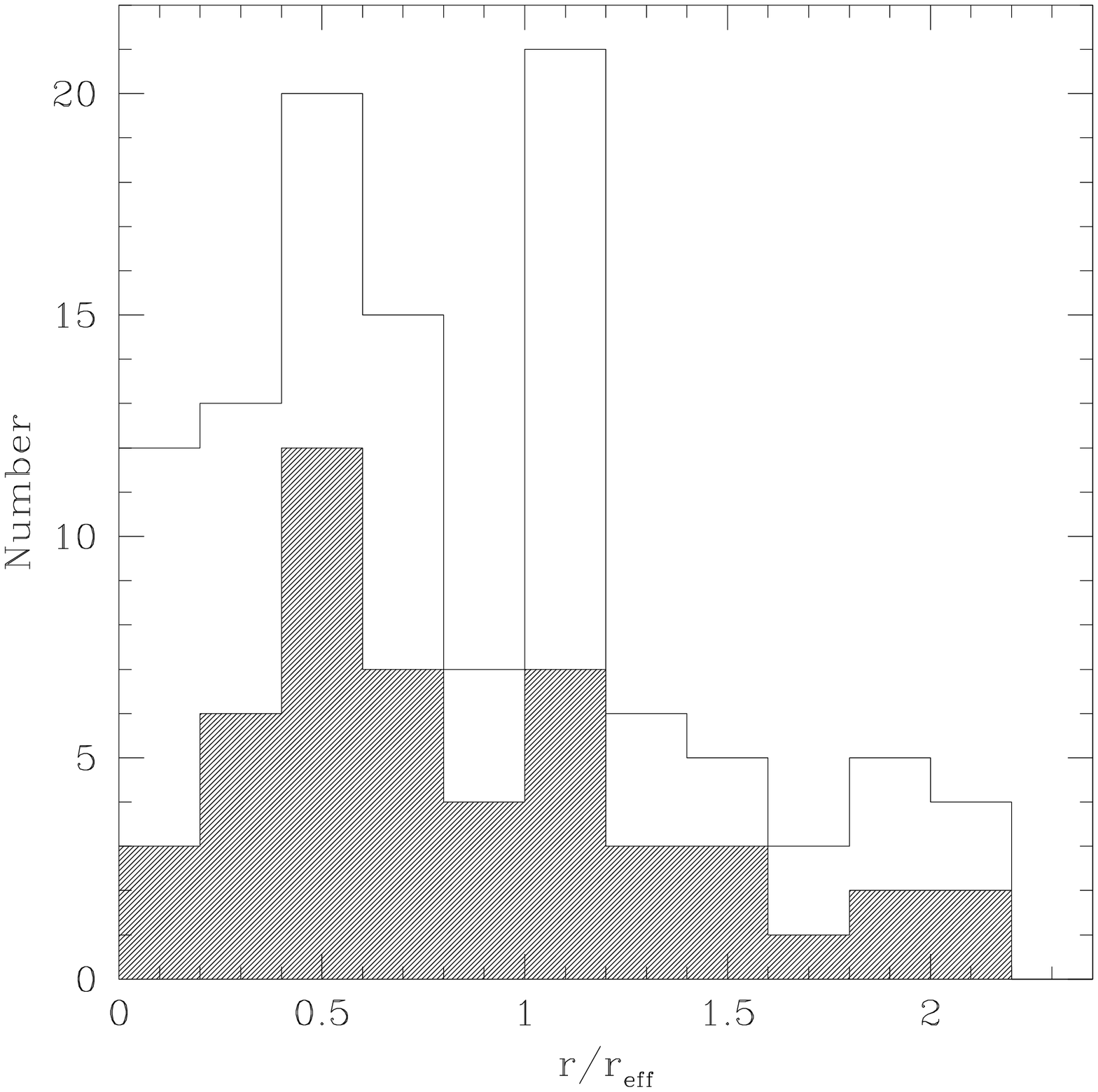}
\caption{Histograms of the number of X-ray sources versus their
projected radius from the center of the galaxy in units of the effective
radius of the galaxy, $r_{\rm eff}$.
The upper histogram is for all of the sources in the $L_X$ complete sample.
The lower shaded histogram shows the X-ray sources identified with GCs.
The histogram bins are $\Delta ( r / r_{\rm eff} ) = 0.2$ wide.
\label{fig:x_deff}}
\end{figure}

We first compare the spatial distributions of GC and non-GC X-ray sources.
Figure~\ref{fig:x_deff} shows histograms of the spatial distributions
of these two types of LMXBs for the $L_X$ complete sample.
There is a slight tendency for the GC LMXBs to be located at larger
radii than the non-GC LMXBs, but it is not very significant.
The median radii of the GC and non-GC LMXBs are 0.65 and 0.49
$r_{\rm eff}$, respectively.
Using the Wilcoxon or Mann-Whitney rank-sum test or the KS test,
the probability that the two samples were drawn from the same
radial distribution can only be rejected at the 93\% level.
In the full sample, this trend is even weaker.
The individual galaxies in our sample have too few sources to allow useful
comparisons.
\citet{kmz02} found that the spatial distributions of GC and
non-GC X-ray sources were similar in NGC~4472.

\subsection{Distribution of GCs} \label{sec:spatial_gc}

\begin{figure}
\epsscale{1.0}
\plotone{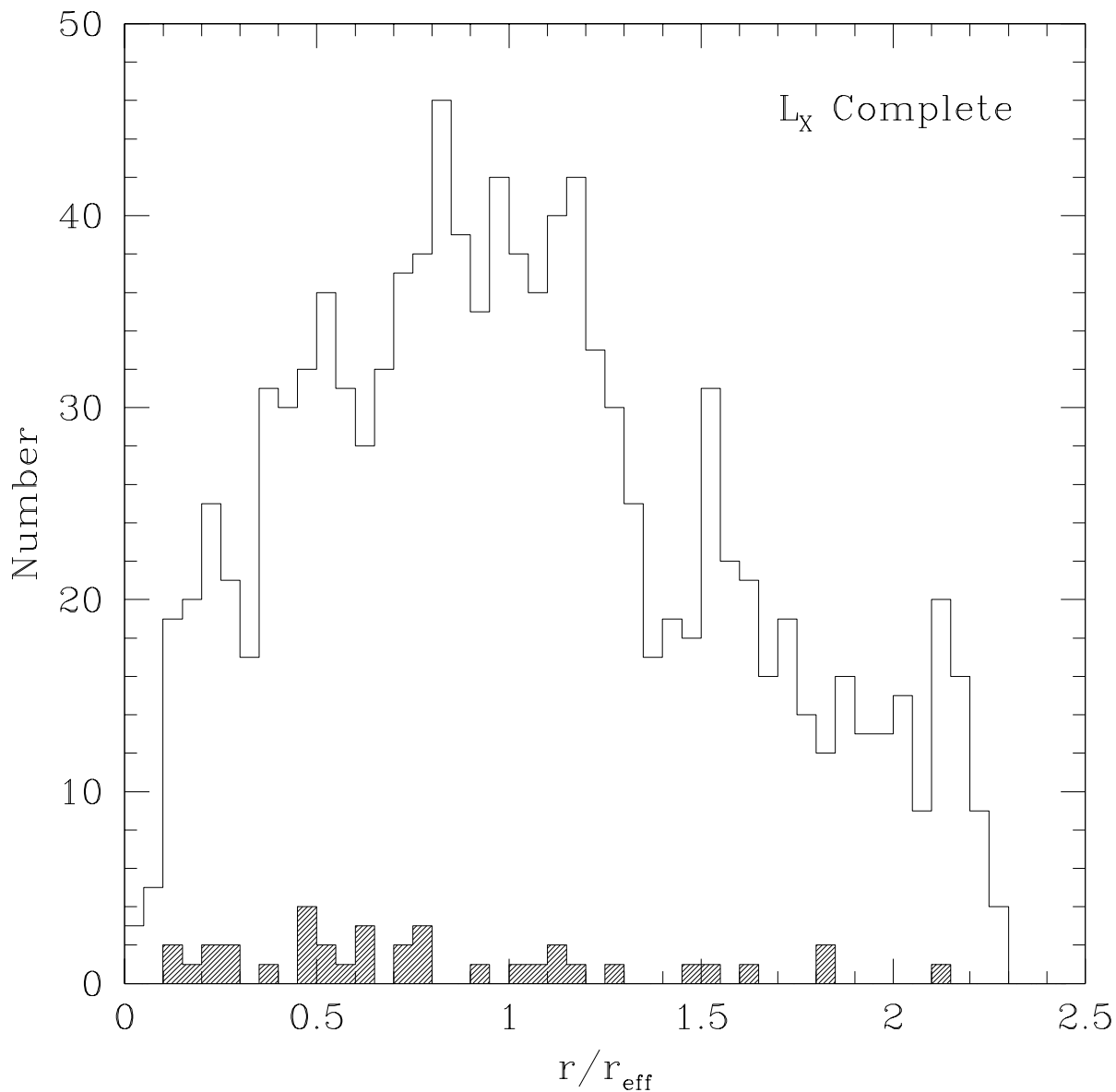}
\caption{Histograms of the number of GCs versus their
projected radius from the center of the galaxy in units of the effective
radius of the galaxy, $r_{\rm eff}$.
The upper histogram is for all of the GCs in our sample.
The lower shaded histogram shows the GCs identified with X-ray sources
in the $L_X$ complete sample.
The histogram bins are $\Delta ( r / r_{\rm eff} ) = 0.05$ wide.
\label{fig:gc_deff}}
\end{figure}

\begin{figure}
\plottwo{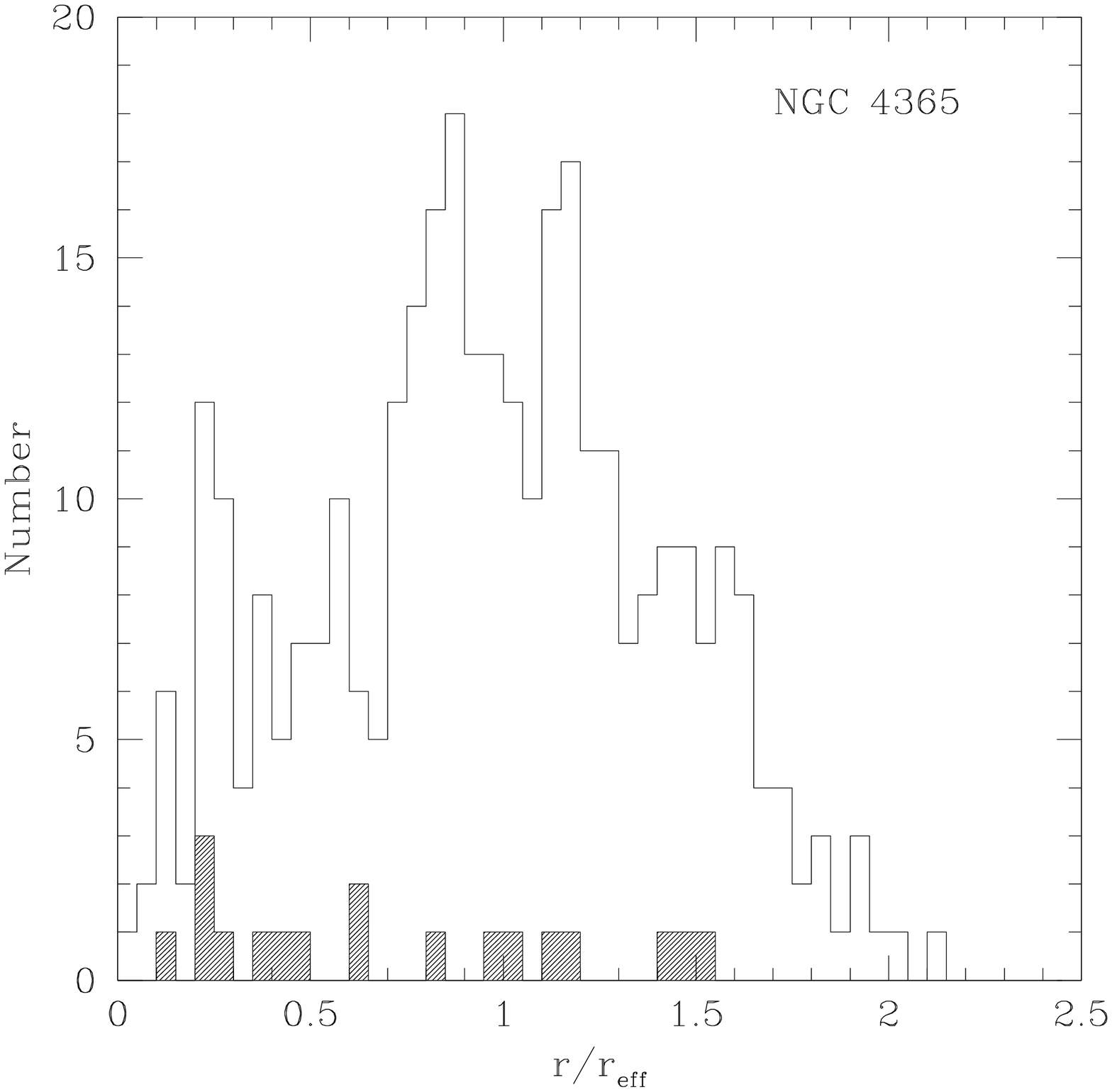}{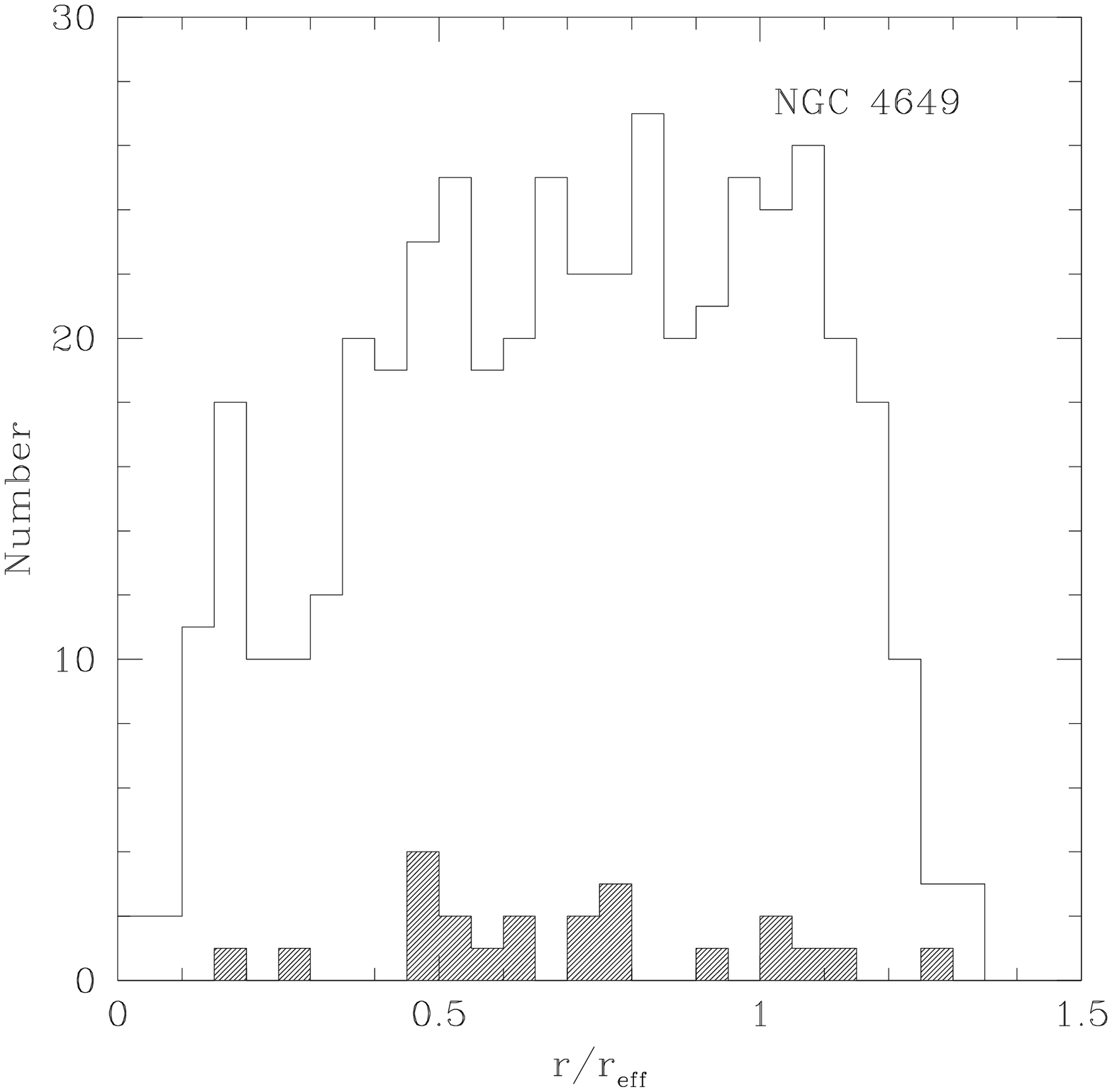}
\caption{Histograms of the number of GCs versus their
projected radius from the center of the galaxy in units of the effective
radius of the galaxy, $r_{\rm eff}$,
for the galaxies NGC~4365 (left) and NGC~4649 (right).
The notation is the same as Figure~\protect\ref{fig:gc_deff}.
\label{fig:gc_deff2}}
\end{figure}

Next, we compared the radial distributions of GCs either containing or
not containing LMXBs.
Figure~\ref{fig:gc_deff} shows histograms of the spatial distribution of
X-ray and non-X-ray GCs in the $L_x$ complete sample.
There appears to be some tendency for the X-ray GCs to be located at
smaller radii.
The median radius of the X-ray GCs is 0.67 $r_{\rm eff}$,
while that for the non-X-ray GCs is 1.00 $r_{\rm eff}$.
The rank-sum test indicates that the radii of the X-ray GCs are smaller,
with a probability that they were drawn from the same distribution of
0.3\%.
Similarly, the KS test finds that the probability that the distributions
are identical is about 0.5\%.
Similar results are found for the full sample, and in each of the
individual galaxies, although the statistics for individual galaxies are
poor.
Figure~\ref{fig:gc_deff2} shows histograms of the spatial distribution of
X-ray and non-X-ray GCs in NGC~4365 and in NGC~4649.
In NGC~4649 and in NGC~4697, the ranges of radii covered (in terms of
$r_{\rm eff}$) are both small, in NGC~4649 because the value
of $r_{\rm eff}$ in large in angular units (Table~\ref{tab:galaxies})
and in NGC~4697 because the GCs are all in an annulus at larger
radii (Figure~\ref{fig:regions}).

Thus, there is at least marginal evidence that the X-ray GCs are more
centrally located than the non-X-ray GCs.
\citet{kmz02} found a similar effect in NGC~4472.
Since the X-ray globular clusters are brighter and redder than the
non-X-ray GCs, this weaker radial trend might be due to these other
stronger correlations, and to a radial trend in the absolute magnitudes or
colors of GCs within our sample.
Indeed, we find that there is a correlation between GC optical luminosity
or color and radius, with the brighter and redder GCs being more likely
to be found at smaller radii.
The optical luminosity gradient is certainly, at least in part, a
selection effect, since it is harder to see faint GC near the center
of the host galaxy.
Other selection effects might be introduced by differing absolute magnitude
limits and differing radial coverage in each galaxy.
In light of various selection effects in the raw data (e.g., the radial
variation of incompleteness in the detection limits of GCs, which differs
from galaxy to galaxy), and the radial variation in the intrinsic properties
of GCs, such as the more spatially concentrated nature of red clusters
with respect to the blue ones seen in most galaxies
\citep*[e.g.,][]{glk96,kws+99},
it is not clear from this data set whether the distance from the
center of a galaxy independently affects the efficiency of LMXB formation
in a GC.
Crude statistical tests with this sample suggest that the radial variations
due to selection effects and intrinsic properties might indeed be able to
reproduce the observed radial gradient.

\section{Conclusions} \label{sec:conclusion}

We have studied the connection between LMXBs and GCs in a sample of
early-type galaxies.
Ultimately, it is hoped that such studies may provide information on
the origin and evolution of LMXBs, and on the evolution of GCs.
In general, the fraction of LMXBs associated with globular
clusters ($f_{\rm X-GC} \approx 10 - 70\%$) is much higher
(by factors of $10^2 - 10^3$) than the fraction of optical light
in GCs.
This result, which is also true for our Galaxy,
indicates that stellar dynamical interactions
in globular clusters are very effective at producing compact binary systems
\citep{cla75,fpr75,hil76}.

There is evidence that the fraction of LMXBs associated with globular
clusters increases along the Hubble
sequence from spiral bulges to S0s to Es to cDs.
The increase is a factor of $\sim$6.
For some time, it has been known that
the specific frequency of globular clusters (the number per unit galaxy
bulge optical luminosity) also increases along the same sequence
\citep[e.g.,][]{har91} by a similar factor.
These two trends might suggest that there are two populations of LMXBs
in early-type galactic systems:
those formed in GCs,
and those formed from field binary stars.
The variation in the fraction of LMXBs in GCs might then result from
the variation in the specific frequency of GCs.
If the number of field LMXBs were proportional to the field optical
luminosity in all galaxies, and the number of GC LMXBs were proportional
to the number of GCs, one might expect the variation in the fraction of
LMXBs in GCs with Hubble type to be considerably smaller than the
increase in the specific frequency of GCs, which is not really the case.
However, variations in the field stellar populations from galaxy to galaxy
might result in variations in the number of field LMXBs per optical
luminosity.
We note that observations suggest that the total number and
X-ray luminosity of LMXBs in galaxies increase in proportion to the
total number of GCs rather than the optical luminosity
\citep{wsk02}.
This suggests that most or all LMXBs are formed in GCs
\citep{gri84,sib01,wsk02},
at least in the earlier type galaxies.
Then, the variation in the fraction seen in GCs today with galaxy type
requires that a higher fraction of LMXBs have escaped from GCs in
later type galaxies (spiral bulges) than earlier galaxies (gEs and cDs).
However, the fraction of LMXBs which would have escaped from GCs in 
later type galaxies would be rather large, and might be inconsistent
with models for the destruction of GCs or the stellar dynamical escape
of LMXBs.
Future tests with larger samples may allow us to determine what fraction of
LMXBs are made in GCs versus the field, and might provide independent
information on the rate of destruction of GCs in galaxies.

The fraction of globular clusters which contain X-ray
sources appears to be roughly constant at $f_{\rm GC-X} \sim 4$\%
in different galaxy types
\citep{kmz02}.
Because of possible incompleteness of the GC samples at the faint end
and because the probability of a GC containing a LMXB correlates with
its optical luminosity, it may be more useful to give the number of
high luminosity LMXBs per optical luminosity of GCs.
In our complete sample, this value is about
$1.5 \times 10^{-7}$ LMXBs per $L_{\odot,I}$ for
$L_X \ga 1 \times 10^{38}$ erg s$^{-1}$ (0.3--10 keV).
The value rises to about
$2.0 \times 10^{-7}$ LMXBs per $L_{\odot,I}$ at lower X-ray luminosities
($L_X \ga 3 \times 10^{37}$ erg s$^{-1}$).
The value for the Milky Way is also very similar to this.
The rate of occurrence of bright LMXBs in GCs
may provide information on the duty cycle of quiescence vs.\
X-ray brightness for GC LMXBs, complementing information from
the detection of quiescent LMXBs in Galactic GCs
\citep[e.g.,][]{ghe+01}.

The strongest trend which we have found is a correlation between
the probability that a GC has an X-ray source and its optical
luminosity.
However, this correlation is consistent with a constant probability of
finding a LMXB per unit optical luminosity;
that is, it seems to result primarily from the larger number of stars
in optically luminous GCs.
Thus, this correlation doesn't provide any direct information on the
physics of the formation and evolution of LMXBs in GCs.
It would be very useful to correlate the probability that a GC contains
a LMXB with the stellar density in the cluster in order to test the role
of stellar dynamical interactions.
The stellar densities could be estimated from the optical magnitudes and
sizes determined from {\it HST} observations.

We find that LMXBs are more likely to be found in redder GCs, and
that this trend is independent of the optical luminosity correlation.
This seems to indicate that the evolution of X-ray binaries in GCs is
affected either by the metallicity or age of the GC, with younger and/or
more metal-rich GCs having more LMXBs.

Models for the evolution of LMXBs suggest that the population should
vary with time after the formation of the stellar population
\citep{wg98,wu01}.
\citet{wg98} argued that the number of LMXBs would initially
rise with time for $\sim$ 1 Gyr, which is set by the time scale for
the secondaries in LMXBs to evolve.
After this, the number of LMXBs should decline slowly with time
as the number of pre-LMXB systems decreases and the existing LMXBs
evolve into millisec pulsar binaries.
Given that the stellar populations in GCs are generally much older than
1 Gyr, we should be in the latter phase of evolution.
Thus, a decline with age would not be unexpected.
However, the arguments in \citet{wg98} apply to LMXBs formed from
the isolated evolution of a population of initial binary stars;
in GCs, new tight binary systems can be produced by stellar dynamical
interactions.
This might reduce (or even reverse) the rate of decline of LMXBs with age.
More detailed models for the evolution of the population of LMXBs in
GCs would be very useful.

The number of LMXBs might also depend on the metallicity of the stellar
population.
A correlation between metallicity and the presence of LMXBs is also seen
in GCs in our Galaxy and in M31
\citep[e.g.,][]{bpf+95}.
This might be due to the fact that higher metallicity stars are larger
\citep{bpf+95};
however, the direct effect of this on the evolution of LMXBs
does not appear to be very strong
\citep[e.g.,][]{rit99}.
Larger stellar sizes might also promote tidal captures to form
LMXBs
\citep{bpf+95}.
Alternatively, metallicity might affect the stellar initial mass function
\citep{gri93} or have some other more complicated effect on the formation
of LMXBs.

There is no strong difference in the X-ray luminosities of GC and non-GC
LMXBs in our sample.
This disagrees with the result found by \citet{alm01} for the galaxy
NGC~1399, where the brightest LMXBs appeared to be positively
correlated with GCs.
On the other hand, our result is consistent with the analysis
of the sources in NGC~4472 by
\citet{kmz02}.
There is a weak tendency for the brightest LMXBs, whose luminosities
exceed the Eddington luminosity for a 1.4 $M_\odot$ neutron star, to
avoid GCs.
That may indicate that black hole X-ray binaries are somewhat less
likely to be found in GCs,
as seems to be true in our Galaxy.
It may be difficult to retain massive BH binaries in GCs
\citep{pzm00}.
In any case, there are some luminous LMXBs associated with GCs.
If one assumes that these most luminous sources contain fairly massive
BHs, it would appear that GCs can retain at least some massive
BH binaries.

There is no clear evidence that the X-ray spectra or variability of GC and
non-GC X-ray sources differ.
A much larger sample and repeated observations with {\it Chandra} would
give stronger limits on the differences in the spectra or variability.

We also find no evidence for a significant difference in the spatial
distribution of GC and non-GC LMXBs.
There is some evidence that X-ray GCs are more likely to be found near
the centers of the galaxies than non-X-ray GCs, but this seems to
be mainly due to an anticorrelation of optical luminosity with radius in
our sample, plus the correlation of LMXBs with the more optically luminous
GCs.
Unfortunately, there are problems with the completeness of both the X-ray
LMXB samples and optical GC samples.
For the X-ray samples, the biggest problem is probably that most of the
galaxies observed by {\it Chandra} so far have been X-ray bright galaxies,
where diffuse emission from hot gas makes it difficult to detect fainter
LMXBs near the centers of the galaxies.
Hopefully, more observations of X-ray faint early-type galaxies will
resolve this problem.
Repeated {\it Chandra} observations would also allow variability to be
detected in more cases, and would allow fainter persistent sources to
be detected in the combined exposure.
The main limitation in the optical samples of GCs is that not all of
the best galaxies have been studied with the {\it HST}, and
the field of view of a single WFPC2 observation is smaller than the angular
size of nearby early-type galaxies.
Multiple observations with the WFPC2 or with the ACS may provide more
complete samples of GCs for more galaxies.

\acknowledgements
We are extremely grateful to JJ Kavelaars for providing his unpublished
list of globular clusters in NGC~4697.
Part of this work was done during a workshop at the Aspen Center for
Physics; J. A. I. and C. L. S. are very grateful to the Center for their
hospitality.
Support for this work was provided by the National Aeronautics and Space
Administration through $Chandra$ Award
Numbers
GO1-2078X,
GO2-3099X,
and
GO2-3100X,
issued by the $Chandra$ X-ray Observatory Center, which is operated by the
Smithsonian Astrophysical Observatory for and on behalf of NASA under
contract NAS8-39073.
Support for E. L. B. was provided by NASA through the {\it Chandra}
Fellowship Program, grant award number PF1-20017, under NASA contract
number
NAS8-39073.
Support for
J. A. I.  was also provided by NASA through the
{\it Chandra} Fellowship Program.
S. W. R. was supported in part by a fellowship from the Virginia Space
Grant Consortium.

\newpage

%
%
\begin{deluxetable}{lcccccccll}
\rotate
\tabletypesize{\small}
\tablewidth{8.2truein}
\tablecaption{Sample Early-Type Galaxies \label{tab:galaxies}}
\tablehead{
\colhead{} &
\colhead{} &
\colhead{$N_H$} &
\colhead{$r_{\rm eff}$} &
\colhead{$D$} &
\colhead{X-ray Exp.} &
\colhead{Count Rate Limit} &
\colhead{$L_X$ Limit} &
\multicolumn{2}{c}{References} \\
\colhead{Galaxy} &
\colhead{Type} &
\colhead{($10^{20}$ cm$^{-2}$)} &
\colhead{(\arcsec)} &
\colhead{(Mpc)} &
\colhead{(ks)} &
\colhead{($10^{-4}$ s$^{-1}$)} &
\colhead{($10^{37}$ erg s$^{-1}$)} &
\colhead{LMXB} &
\colhead{GC} \\
}
\startdata
NGC~1553 & S0 & 1.50 & 63 & 18.5 & 23.2 & 4.3 &    12.3  & BSI & KWb \\
NGC~4365 & E3 & 1.63 & 50 & 20.4 & 40.4 & 2.7 &    10.6  & SSI & KWa \\
NGC~4649 & E2 & 2.13 & 82 & 16.8 & 36.8 & 3.1 & \phn7.0  & RSI & KWa \\
NGC~4697 & E6 & 2.14 & 72 & 11.7 & 39.4 & 2.6 & \phn2.7  & SIB & K   \\
\enddata
\tablerefs{
BSI \citep*{bsi01};
K (J. Kavelaars 2000, private communication);
KWa \citep{kw01a};
KWb \citep{kw01b};
RSI \citep{rsi03};
SIB \citep{sib00,sib01};
SSI \citep*{ssi03}
}
\end{deluxetable}

\newpage

%
%
\begin{deluxetable}{lcccccccccc}
\tabletypesize{\scriptsize}
\tablecaption{X-rays Sources in GC Survey Regions \label{tab:src}}
\tablehead{
\colhead{} &
\colhead{Src.} &
\colhead{Count Rate} &
\colhead{$L_X$ (0.3--10 keV)} &
\colhead{} &
\colhead{} &
\colhead{} &
\colhead{} &
\colhead{$M_I$} &
\colhead{$V-I$} &
\colhead{$L_X$} \\
\colhead{Galaxy} &
\colhead{No.} &
\colhead{($10^{-4}$ s$^{-1}$)} &
\colhead{($10^{37}$ erg s$^{-1}$)} &
\colhead{H21$^0$} &
\colhead{H31$^0$} &
\colhead{Var.?} &
\colhead{GC?} &
\colhead{(mag)} &
\colhead{(mag)} &
\colhead{Complete?} \\
}
\startdata
NGC~1553&\phn3&   15.07$\pm$2.67&   \phn43.3&$-$0.34&$-$0.35& &       &   $\cdots$&$\cdots$ & $\surd$ \\
NGC~1553&\phn4&\phn7.64$\pm$1.94&   \phn21.9&$-$0.40&$-$0.47& &$\surd$&   $-$10.63&   1.054 & $\surd$ \\
NGC~1553&\phn5&\phn9.04$\pm$2.08&   \phn25.9&$-$0.20&$-$0.13& &       &   $\cdots$&$\cdots$ & $\surd$ \\
NGC~1553&\phn7&\phn5.47$\pm$1.62&   \phn15.7&$-$0.66&$-$0.11& &       &   $\cdots$&$\cdots$ & $\surd$ \\
NGC~1553&\phn8&\phn8.46$\pm$2.03&   \phn24.3&$-$0.14&$-$0.59& &       &   $\cdots$&$\cdots$ & $\surd$ \\
NGC~1553&   11&\phn4.56$\pm$1.50&   \phn13.1&$+$0.09&$-$0.69& &       &   $\cdots$&$\cdots$ & $\surd$ \\
NGC~1553&   16&   40.04$\pm$4.22&      114.9&$-$0.10&$-$0.54& &       &   $\cdots$&$\cdots$ & $\surd$ \\
NGC~1553&   17&   66.51$\pm$5.41&      190.9&$-$0.15&$-$0.65&V&       &   $\cdots$&$\cdots$ & $\surd$ \\
NGC~1553&   18&\phn7.24$\pm$1.83&   \phn20.8&$-$0.09&$-$0.51& &       &   $\cdots$&$\cdots$ & $\surd$ \\
NGC~1553&   19&\phn4.46$\pm$1.52&   \phn12.8&$-$0.57&$-$0.67& &       &   $\cdots$&$\cdots$ & $\surd$ \\
NGC~1553&   23&\phn6.14$\pm$1.68&   \phn17.6&$-$0.43&$-$0.72& &$\surd$&   $-$12.17&   1.088 & $\surd$ \\
NGC~4365&\phn1&   37.34$\pm$3.18&      152.5&$-$0.17&$-$0.53& &       &   $\cdots$&$\cdots$ & $\surd$ \\
NGC~4365&\phn2&   14.23$\pm$1.98&   \phn58.1&$-$0.30&$-$0.34& &       &   $\cdots$&$\cdots$ & $\surd$ \\
NGC~4365&\phn3&   48.89$\pm$3.61&      199.6&$-$0.13&$-$0.40& &       &   $\cdots$&$\cdots$ & $\surd$ \\
NGC~4365&\phn4&\phn8.86$\pm$1.56&   \phn36.2&$+$0.18&$+$0.06& &       &   $\cdots$&$\cdots$ & $\surd$ \\
NGC~4365&\phn5&\phn4.59$\pm$1.16&   \phn18.7&$-$0.42&$-$0.42& &$\surd$&\phn$-$9.26&   0.936 & $\surd$ \\
NGC~4365&\phn6&\phn5.94$\pm$1.35&   \phn24.3&$+$0.67&$-$0.77& &       &   $\cdots$&$\cdots$ & $\surd$ \\
NGC~4365&\phn7&   13.66$\pm$1.97&   \phn55.8&$-$0.26&$-$0.23& &       &   $\cdots$&$\cdots$ & $\surd$ \\
NGC~4365&\phn8&\phn3.37$\pm$1.01&   \phn13.8&$-$0.13&$-$0.28& &$\surd$&   $-$11.18&   1.069 & $\surd$ \\
NGC~4365&\phn9&\phn7.11$\pm$1.43&   \phn29.0&$+$0.31&$-$0.11& &$\surd$&\phn$-$9.24&   1.038 & $\surd$ \\
NGC~4365&   10&\phn5.40$\pm$1.28&   \phn22.0&$+$0.44&$-$0.06& &$\surd$&\phn$-$8.51&   1.056 & $\surd$ \\
NGC~4365&   11&\phn3.53$\pm$1.04&   \phn14.4&$+$0.01&$-$0.41& &       &   $\cdots$&$\cdots$ & $\surd$ \\
NGC~4365&   12&\phn7.18$\pm$1.45&   \phn29.3&$+$0.22&$-$0.25& &$\surd$&   $-$10.14&   1.241 & $\surd$ \\
NGC~4365&   13&\phn4.16$\pm$1.14&   \phn17.0&$-$0.76&$-$0.43& &$\surd$&\phn$-$9.95&   0.926 & $\surd$ \\
NGC~4365&   16&\phn5.74$\pm$1.30&   \phn23.5&$-$0.11&$-$0.42&V&       &   $\cdots$&$\cdots$ & $\surd$ \\
NGC~4365&   17&\phn3.22$\pm$0.97&   \phn13.1&$+$0.58&$-$0.26& &$\surd$&   $-$10.68&   1.071 & $\surd$ \\
NGC~4365&   18&   11.78$\pm$1.80&   \phn48.1&$+$0.05&$-$0.28& &$\surd$&   $-$11.37&   1.125 & $\surd$ \\
NGC~4365&   22&\phn3.93$\pm$1.06&   \phn16.0&$+$0.24&$+$0.02& &       &   $\cdots$&$\cdots$ & $\surd$ \\
NGC~4365&   25&\phn6.86$\pm$1.38&   \phn28.0&$+$0.20&$-$0.06& &$\surd$&   $-$11.18&   1.057 & $\surd$ \\
NGC~4365&   26&\phn2.69$\pm$0.89&   \phn11.0&$+$0.33&$+$0.00& &$\surd$&\phn$-$8.64&   1.226 & \\
NGC~4365&   29&\phn2.65$\pm$0.89&   \phn10.8&$+$0.24&$-$0.01&V&       &   $\cdots$&$\cdots$ & \\
NGC~4365&   30&\phn6.67$\pm$1.36&   \phn27.2&$-$0.37&$-$0.49& &       &   $\cdots$&$\cdots$ & $\surd$ \\
NGC~4365&   31&\phn2.66$\pm$0.89&   \phn10.9&$-$0.05&$-$0.88&V&       &   $\cdots$&$\cdots$ & \\
NGC~4365&   32&\phn4.05$\pm$1.08&   \phn16.5&$+$0.26&$+$0.02& &       &   $\cdots$&$\cdots$ & $\surd$ \\
NGC~4365&   33&\phn5.03$\pm$1.51&   \phn20.5&$+$0.41&$-$0.76& &       &   $\cdots$&$\cdots$ & $\surd$ \\
NGC~4365&   34&\phn2.94$\pm$0.93&   \phn12.0&$-$0.59&$-$0.71& &$\surd$&   $-$10.00&   1.205 & \\
NGC~4365&   37&\phn2.79$\pm$0.92&   \phn11.4&$-$0.47&$-$0.48& &$\surd$&   $-$10.13&   1.290 & \\
NGC~4365&   38&   18.65$\pm$2.27&   \phn76.1&$-$0.38&$-$0.92& &       &   $\cdots$&$\cdots$ & $\surd$ \\
NGC~4365&   40&\phn3.60$\pm$1.03&   \phn14.7&$-$0.07&$-$0.85& &$\surd$&   $-$11.38&   1.101 & $\surd$ \\
NGC~4365&   42&   11.86$\pm$1.79&   \phn48.4&$+$0.08&$-$0.22& &       &   $\cdots$&$\cdots$ & $\surd$ \\
NGC~4365&   43&\phn3.95$\pm$1.06&   \phn16.1&$+$0.46&$+$0.18& &       &   $\cdots$&$\cdots$ & $\surd$ \\
NGC~4365&   44&\phn3.22$\pm$1.07&   \phn13.1&$-$0.05&$-$0.70& &$\surd$&   $-$11.15&   1.172 & $\surd$ \\
NGC~4365&   45&\phn4.21$\pm$1.19&   \phn17.2&$+$0.19&$+$0.09& &$\surd$&   $-$10.98&   0.958 & $\surd$ \\
NGC~4365&   46&\phn2.62$\pm$0.86&   \phn10.7&$-$0.05&$-$0.70& &       &   $\cdots$&$\cdots$ & \\
NGC~4365&   47&   10.72$\pm$1.68&   \phn43.8&$+$0.04&$-$0.63& &       &   $\cdots$&$\cdots$ & $\surd$ \\
NGC~4365&   48&\phn2.70$\pm$0.89&   \phn11.0&$+$0.15&$+$0.00&V&$\surd$&\phn$-$9.86&   1.180 & \\
NGC~4365&   50&\phn6.29$\pm$1.31&   \phn25.7&$-$0.26&$-$0.72& &$\surd$&\phn$-$9.69&   1.120 & $\surd$ \\
NGC~4365&   53&\phn5.85$\pm$1.26&   \phn23.9&$+$0.55&$+$0.44& &$\surd$&   $-$10.32&   1.027 & $\surd$ \\
NGC~4649&\phn2&   33.52$\pm$3.46&   \phn75.4&$-$0.61&$-$0.90& &       &   $\cdots$&$\cdots$ & $\surd$ \\
NGC~4649&\phn3&   20.43$\pm$2.90&   \phn45.9&$-$0.32&$-$0.47& &$\surd$&\phn$-$9.93&   1.081 & $\surd$ \\
NGC~4649&\phn4&   20.80$\pm$2.92&   \phn46.8&$-$0.44&$-$0.40& &       &   $\cdots$&$\cdots$ & $\surd$ \\
NGC~4649&\phn5&   13.11$\pm$2.50&   \phn29.5&$-$0.22&$-$0.68& &       &   $\cdots$&$\cdots$ & $\surd$ \\
NGC~4649&\phn6&   21.49$\pm$3.01&   \phn48.3&$+$0.13&$-$0.55& &$\surd$&   $-$11.19&   0.999 & $\surd$ \\
NGC~4649&\phn9&   15.94$\pm$2.57&   \phn35.8&$-$0.21&$-$0.42& &       &   $\cdots$&$\cdots$ & $\surd$ \\
NGC~4649&   12&\phn9.94$\pm$1.98&   \phn22.4&$+$0.09&$-$0.23& &       &   $\cdots$&$\cdots$ & $\surd$ \\
NGC~4649&   16&\phn4.06$\pm$1.28&\phn\phn9.1&$-$1.00&$-$0.28& &       &   $\cdots$&$\cdots$ & \\
NGC~4649&   18&   20.77$\pm$2.80&   \phn46.7&$-$0.06&$-$0.23& &$\surd$&   $-$11.03&   1.116 & $\surd$ \\
NGC~4649&   19&\phn9.43$\pm$2.00&   \phn21.2&$-$0.11&$-$1.00& &       &   $\cdots$&$\cdots$ & $\surd$ \\
NGC~4649&   20&   15.20$\pm$2.36&   \phn34.2&$-$0.08&$-$0.22& &$\surd$&   $-$10.22&   1.225 & $\surd$ \\
NGC~4649&   21&\phn6.38$\pm$1.69&   \phn14.3&$-$0.15&$-$0.34& &       &   $\cdots$&$\cdots$ & $\surd$ \\
NGC~4649&   22&\phn5.15$\pm$1.43&   \phn11.6&$+$0.01&$-$0.56& &$\surd$&   $-$11.23&   1.246 & \\
NGC~4649&   23&\phn9.29$\pm$1.85&   \phn20.9&$+$1.00&$+$1.00& &$\surd$&   $-$11.27&   1.149 & $\surd$ \\
NGC~4649&   25&   16.21$\pm$2.36&   \phn36.5&$+$0.20&$-$0.10& &       &   $\cdots$&$\cdots$ & $\surd$ \\
NGC~4649&   26&\phn6.69$\pm$1.66&   \phn15.1&$+$0.21&$+$0.03& &       &   $\cdots$&$\cdots$ & $\surd$ \\
NGC~4649&   27&\phn7.68$\pm$1.75&   \phn17.3&$-$0.14&$-$0.57& &       &   $\cdots$&$\cdots$ & $\surd$ \\
NGC~4649&   28&\phn7.39$\pm$1.72&   \phn16.6&$-$0.44&$-$0.11& &$\surd$&   $-$10.19&   1.085 & $\surd$ \\
NGC~4649&   29&   11.72$\pm$2.10&   \phn26.4&$+$0.05&$-$0.68& &$\surd$&   $-$10.28&   1.228 & $\surd$ \\
NGC~4649&   31&   18.10$\pm$2.49&   \phn40.7&$+$0.00&$-$0.55& &$\surd$&   $-$10.57&   1.220 & $\surd$ \\
NGC~4649&   32&\phn7.53$\pm$1.77&   \phn16.9&$-$0.50&$-$0.54& &$\surd$&   $-$10.41&   1.031 & $\surd$ \\
NGC~4649&   33&\phn6.16$\pm$1.60&   \phn13.8&$-$0.42&$-$0.82& &$\surd$&\phn$-$8.46&   1.335 & $\surd$ \\
NGC~4649&   40&\phn5.23$\pm$1.45&   \phn11.8&$+$0.09&$-$0.01& &       &   $\cdots$&$\cdots$ & \\
NGC~4649&   41&\phn5.92$\pm$1.48&   \phn13.3&$+$1.00&$+$1.00& &       &   $\cdots$&$\cdots$ & $\surd$ \\
NGC~4649&   42&   11.38$\pm$2.01&   \phn25.6&$+$0.09&$-$0.40& &$\surd$&   $-$10.19&   1.217 & $\surd$ \\
NGC~4649&   45&\phn6.42$\pm$1.52&   \phn14.4&$+$0.04&$-$0.43& &$\surd$&\phn$-$8.95&   1.218 & $\surd$ \\
NGC~4649&   46&\phn6.96$\pm$1.63&   \phn15.7&$+$0.54&$+$0.19& &$\surd$&\phn$-$9.50&   1.185 & $\surd$ \\
NGC~4649&   47&\phn5.88$\pm$1.51&   \phn13.2&$-$0.70&$-$0.54& &$\surd$&   $-$10.19&   1.282 & $\surd$ \\
NGC~4649&   48&\phn8.74$\pm$1.72&   \phn19.7&$+$0.03&$-$0.04& &$\surd$&   $-$10.60&   1.218 & $\surd$ \\
NGC~4649&   49&\phn4.89$\pm$1.36&   \phn11.0&$+$0.44&$-$0.71& &       &   $\cdots$&$\cdots$ & \\
NGC~4649&   53&   10.12$\pm$1.94&   \phn22.8&$-$0.04&$-$0.06& &$\surd$&   $-$10.21&   1.320 & $\surd$ \\
NGC~4649&   57&\phn3.89$\pm$1.16&\phn\phn8.7&$+$0.03&$-$0.53& &$\surd$&   $-$11.33&   1.173 & \\
NGC~4649&   58&\phn9.33$\pm$1.76&   \phn21.0&$-$0.20&$-$0.32& &       &   $\cdots$&$\cdots$ & $\surd$ \\
NGC~4649&   61&\phn3.85$\pm$1.16&\phn\phn8.6&$+$0.31&$-$0.86& &       &   $\cdots$&$\cdots$ & \\
NGC~4649&   62&\phn4.41$\pm$1.21&\phn\phn9.9&$+$1.00&$+$1.00& &$\surd$&\phn$-$9.77&   1.058 & \\
NGC~4649&   63&\phn9.51$\pm$1.74&   \phn21.4&$-$0.48&$-$0.47& &$\surd$&   $-$11.46&   1.209 & $\surd$ \\
NGC~4649&   65&\phn3.75$\pm$1.18&\phn\phn8.4&$-$0.90&$-$0.64& &       &   $\cdots$&$\cdots$ & \\
NGC~4649&   68&\phn4.87$\pm$1.28&   \phn10.9&$-$0.28&$-$0.44& &       &   $\cdots$&$\cdots$ & \\
NGC~4649&   69&   38.18$\pm$3.38&   \phn85.8&$-$0.20&$-$0.69& &       &   $\cdots$&$\cdots$ & $\surd$ \\
NGC~4649&   70&   19.65$\pm$2.44&   \phn44.2&$+$0.11&$-$0.50& &$\surd$&\phn$-$8.24&   1.227 & $\surd$ \\
NGC~4649&   71&\phn4.21$\pm$1.25&\phn\phn9.5&$+$0.53&$+$0.20& &       &   $\cdots$&$\cdots$ & \\
NGC~4649&   72&\phn6.34$\pm$1.52&   \phn14.3&$-$0.29&$-$0.38& &       &   $\cdots$&$\cdots$ & $\surd$ \\
NGC~4649&   73&\phn3.35$\pm$1.11&\phn\phn7.5&$+$0.74&$+$0.17& &       &   $\cdots$&$\cdots$ & \\
NGC~4649&   74&   16.57$\pm$2.26&   \phn37.3&$-$0.09&$-$0.18& &       &   $\cdots$&$\cdots$ & $\surd$ \\
NGC~4649&   75&   41.29$\pm$3.53&   \phn92.8&$-$0.12&$-$0.44& &       &   $\cdots$&$\cdots$ & $\surd$ \\
NGC~4649&   80&   10.88$\pm$1.88&   \phn24.5&$-$0.01&$-$0.58& &$\surd$&   $-$10.29&   0.964 & $\surd$ \\
NGC~4649&   85&   17.74$\pm$2.37&   \phn39.9&$+$0.01&$-$0.59& &       &   $\cdots$&$\cdots$ & $\surd$ \\
NGC~4697&   53&   25.75$\pm$2.62&   \phn26.8&$-$0.17&$-$0.54& &$\surd$&\phn$-$7.40&$\cdots$ & $\surd$ \\
NGC~4697&   54&\phn3.61$\pm$1.10&\phn\phn3.7&$-$0.47&$-$0.48& &$\surd$&   $-$11.23&$\cdots$ & \\
NGC~4697&   55&\phn2.70$\pm$0.88&\phn\phn2.8&$+$0.43&$+$0.39&V&$\surd$&\phn$-$8.16&$\cdots$ & \\
NGC~4697&   57&   10.16$\pm$1.64&   \phn10.6&$-$0.04&$-$0.34& &       &   $\cdots$&$\cdots$ & \\
NGC~4697&   58&\phn4.21$\pm$1.08&\phn\phn4.4&$+$0.29&$-$0.13&V&       &   $\cdots$&$\cdots$ & \\
NGC~4697&   59&\phn3.87$\pm$1.02&\phn\phn4.0&$-$0.07&$-$0.75& &       &   $\cdots$&$\cdots$ & \\
NGC~4697&   60&\phn2.62$\pm$0.86&\phn\phn2.7&$-$0.19&$-$0.13& &       &   $\cdots$&$\cdots$ & \\
NGC~4697&   61&   13.68$\pm$1.89&   \phn14.2&$-$0.50&$-$0.62&V&$\surd$&\phn$-$7.91&$\cdots$ & $\surd$ \\
NGC~4697&   62&\phn8.56$\pm$1.64&\phn\phn8.9&$-$0.07&$-$0.28& &       &   $\cdots$&$\cdots$ & \\
NGC~4697&   63&\phn7.46$\pm$1.51&\phn\phn7.7&$-$0.44&$-$0.26& &       &   $\cdots$&$\cdots$ & \\
NGC~4697&   64&   17.34$\pm$2.24&   \phn18.0&$-$0.12&$-$0.93&V&$\surd$&\phn$-$9.37&$\cdots$ & $\surd$ \\
NGC~4697&   65&   26.98$\pm$2.68&   \phn28.0&$+$0.18&$-$0.37& &       &   $\cdots$&$\cdots$ & $\surd$ \\
NGC~4697&   66&\phn7.86$\pm$1.44&\phn\phn8.2&$+$0.00&$-$0.86& &       &   $\cdots$&$\cdots$ & \\
NGC~4697&   67&\phn3.67$\pm$1.10&\phn\phn3.8&$+$0.03&$+$0.32& &$\surd$&\phn$-$6.59&$\cdots$ & \\
NGC~4697&   68&\phn3.27$\pm$1.02&\phn\phn3.4&$+$0.23&$-$1.00& &       &   $\cdots$&$\cdots$ & \\
NGC~4697&   69&   23.51$\pm$2.59&   \phn24.4&$-$0.00&$-$0.29& &$\surd$&\phn$-$7.50&$\cdots$ & $\surd$ \\
\enddata
\end{deluxetable}

\newpage

%
%
\begin{deluxetable}{lccccccccccccc}
\rotate
\tabletypesize{\footnotesize}
\tablewidth{8.7truein}
\tablecaption{Source Statistics in Sample Galaxies \label{tab:statistics}}
\tablehead{
 &
 &
 &
\multicolumn{5}{c}{$L_X$ Complete Sample} &
&
\multicolumn{5}{c}{Full Sample} \\
\cline{4-8}
\cline{10-14} & \\
\colhead{} &
\colhead{} &
\colhead{} &
\colhead{} &
\colhead{} &
\colhead{$f_{\rm X-GC}$} &
\colhead{$f_{\rm GC-X}$} &
\colhead{$N_{\rm Both}/L_I$} &
&
\colhead{} &
\colhead{} &
\colhead{$f_{\rm X-GC}$} &
\colhead{$f_{\rm GC-X}$} &
\colhead{$N_{\rm Both}/L_I$} \\
\colhead{Galaxy} &
\colhead{Type} &
\colhead{$N_{\rm GC}$} &
\colhead{$N_{\rm LMXB}$} &
\colhead{$N_{\rm Both}$} &
\colhead{(\%)} &
\colhead{(\%)} &
\colhead{($10^{-7} \, L_{\odot,I}^{-1}$)} &
&
\colhead{$N_{\rm LMXB}$} &
\colhead{$N_{\rm Both}$} &
\colhead{(\%)} &
\colhead{(\%)} &
\colhead{($10^{-7} \, L_{\odot,I}^{-1}$)} \\
}
\startdata
NGC~1553 & S0 & \phn70 &    11 & \phn2 & $18^{+9}_{-12}$ & $2.9^{+1.7}_{-2.0}$
& 1.0$\pm$0.7 &&  11 & \phn2 & $18^{+9}_{-12}$ & $2.9^{+1.7}_{-2.0}$ &
1.0$\pm$0.7 \\
NGC~4365 & E3 &    325 &    30 &    14 & $47^{+7}_{-12}$ & $4.3^{+1.1}_{-1.1}$
& 1.9$\pm$0.5 &&  37 &    18 & $49^{+6}_{-10}$ & $5.5^{+1.3}_{-1.3}$ &
2.0$\pm$0.5 \\
NGC~4649 & E2 &    445 &    36 &    19 & $53^{+7}_{-10}$ & $4.3^{+0.9}_{-1.0}$
& 1.7$\pm$0.4 &&  47 &    22 & $47^{+6}_{-9}$ & $4.9^{+1.0}_{-1.1}$ &
2.0$\pm$0.4 \\
NGC~4697 & E6 &    263 & \phn5 & \phn4 & $80^{+5}_{-34}$ & $1.5^{+0.7}_{-0.7}$
& 1.6$\pm$0.8 &&  16 & \phn7 & $44^{+9}_{-16}$ & $2.7^{+0.9}_{-1.1}$ &
2.8$\pm$1.1 \\
\enddata
\end{deluxetable}

\newpage

\begin{figure}
\plotone{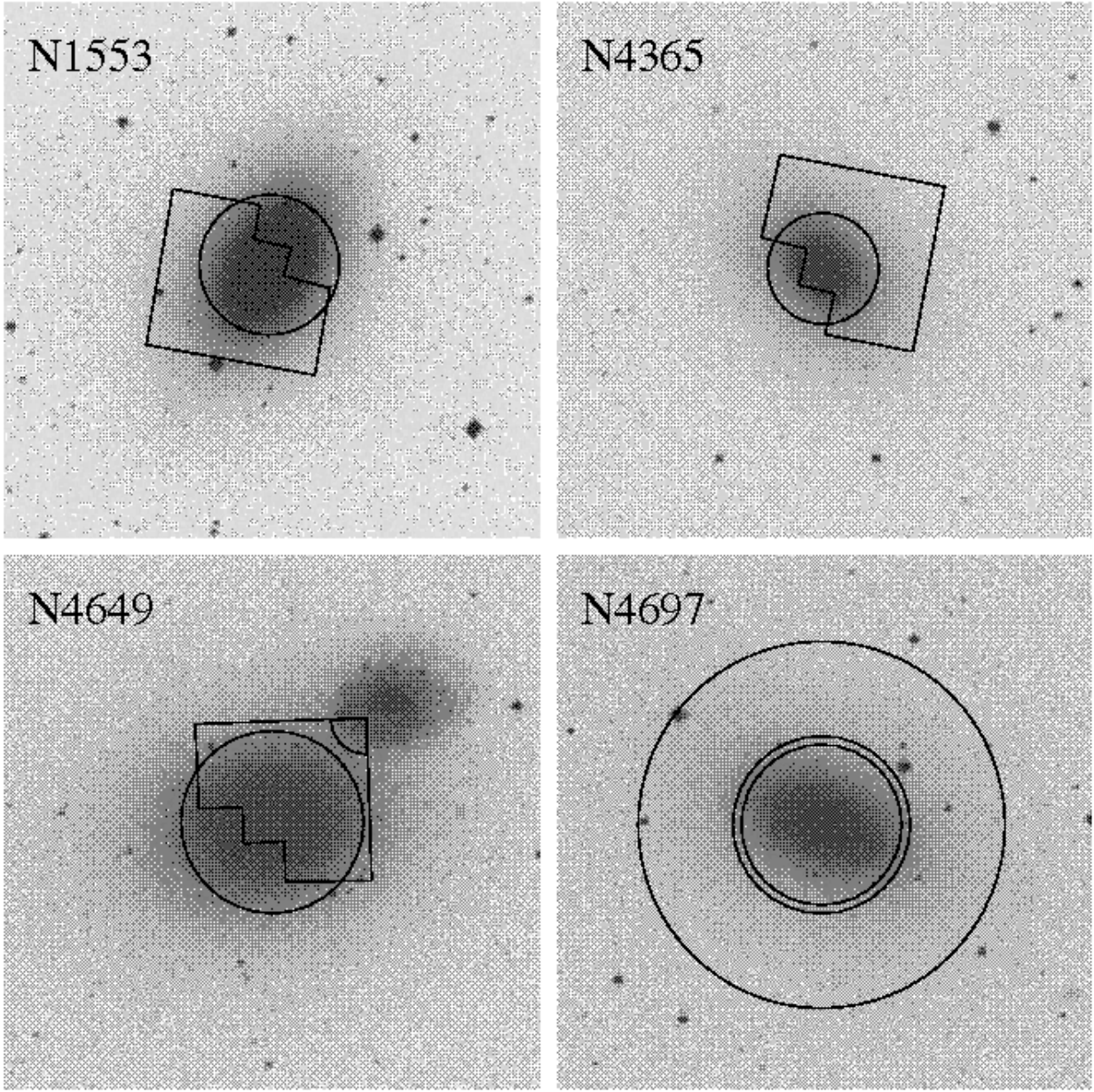}
\caption{Regions covered by the GC survey for each of the sample galaxies
are supposed on Digital Sky Survey (DSS) images of the galaxies.
The images are $8\arcmin \times 8\arcmin$, which is roughly the field of
view of the Chandra images used to detect sources.
In each case, the inner circle has a radius of one effective radius.
For NGC~1553, NGC~4365, and NGC~4649, the polygonal region is approximate
area of the {\it HST} GC survey.
In NGC~4649, the quarter-circle region to the northwest was excluded as
it appeared to mainly contain objects associated with the companion
spiral NGC~4647.
For NGC~4697, the GC survey region is the annulus between the two
outer circles.
\label{fig:regions}}
\end{figure}

\newpage

\begin{thebibliography}{}

\bibitem[Angelini et al.(2001)Angelini, Loewenstein, \& Mushotzky]{alm01}
Angelini, L., Loewenstein, M., \& Mushotzky, R.~F. 2001, ApJ, 557, L35

\bibitem[Barmby \& Huchra(2001)]{bh01}
Barmby, P., \& Huchra, J. P. 2001, AJ, 119, 2349

\bibitem[Bellazzini et al.(1995)]{bpf+95}
Bellazzini, M., Pasquali, A., Federici, L., Ferraro, F.~R., \&
Pecci, F.~F. 1995, ApJ, 439, 687

\bibitem[Blanton et al.(2001)Blanton, Sarazin, \& Irwin]{bsi01}
Blanton, E. L., Sarazin, C. L., \& Irwin, J. A. 2001, ApJ, 552, 106 (BSI)

\bibitem[Clark(1975)]{cla75}
Clark, G. W. 1975, ApJ, 199, L143

\bibitem[{{Cutri} {et~al.}(2001)}]{Cut+01}
{Cutri}, R.~M., {et~al.} 2001, {Explanatory Supplement to the 2MASS Second
  Incremental Data Release}, {The Two Micron All Sky Survey at IPAC
Website:
  http://www.ipac.caltech.edu/2mass/releases/second/doc/explsup.html}

\bibitem[Dickey \& Lockman(1990)]{dl90}
Dickey, J. M., \& Lockman, F. J. 1990, ARA\&A, 28, 215

\bibitem[Di Stefano et al.(2002)]{dkg+02}
Di Stefano, R., Kong, A. K., Garcia, M. R., Barmby, P., Greiner, J.,
Murray, S. S., \& Primini, F. A. 2002, ApJ, 570, 618

\bibitem[Fabian et al.(1975)Fabian, Pringle, \& Rees]{fpr75}
Fabian, A. C., Pringle, J. E., \& Rees, M. J. 1975, MNRAS, 172, 15

\bibitem[Forman et al.(1985)Forman, Jones, \& Tucker]{fjt85}
Forman, W., Jones, C., \& Tucker W. C. 1985, ApJ, 293, 102

\bibitem[Geisler et al.(1996)Geisler, Lee, \& Kim]{glk96}
Geisler, D., Lee, M. G., \& Kim, E. 1996, AJ, 111, 1529

\bibitem[Grindlay(1984)]{gri84}
Grindlay, J. E. 1984, AdSpR, 3, 19

\bibitem[Grindlay(1993)]{gri93}
Grindlay, J. E. 1993, in ASP Conf.\ Ser.\ 48: The Globular Cluster-Galaxy
Connection, ed.\ G. H. Smith \& J. P. Brodie (San Francisco: ASP), 156

\bibitem[Grindlay et al.(2001)]{ghe+01}
Grindlay, J. E., Heinke, C., Edmonds, P. D., \& Murray, S. S.
2001, Science, 292, 2290

\bibitem[Hanes(1977)]{han77}
Hanes, D. A. 1977, MNRAS, 84, 45

\bibitem[Harris(1991)]{har91}
Harris, W. E. 1991, ARAA, 29, 543

\bibitem[Harris(1996)]{har96}
Harris, W. E. 1996, AJ, 112, 1487

\bibitem[Hertz \& Grindlay(1983)]{hg83}
Hertz, P., \& Grindlay, J. E. 1983, ApJ, 275, 105

\bibitem[Hills(1976)]{hil76}
Hills, J. G. 1976, MNRAS, 175, 1

\bibitem[Kahabka \& van den Heuvel(1997)]{kvdh97}
Kahabka, P., \& van den Heuvel E. P. J. 1997, ARAA 35, 69

\bibitem[Katz(1975)]{kat75}
Katz, J. I. 1975, Nature, 253, 698

\bibitem[Kavelaars(2000)]{kav00}
Kavelaars. J. J. 2000, private communication

\bibitem[Kraft et al.(2000)]{kra+00}
Kraft, R. P., et al.\ 2000, ApJ, 531, L9

\bibitem[Kundu et al.(1999)]{kws+99}
Kundu, A., Whitmore, B. C., Sparks, W. B., Macchetto, F. D., Zepf, S. E.,
\& Ashman, K. M. 1999, ApJ, 513, 733

\bibitem[Kundu \& Whitmore(2001a)]{kw01a}
Kundu, A., \& Whitmore, B. C. 2001a, AJ, 121, 2950 (KWa)

\bibitem[Kundu \& Whitmore(2001b)]{kw01b}
Kundu, A., \& Whitmore, B. C. 2001b, AJ, 122, 1251 (KWb)

\bibitem[Kundu et al.(2002)Kundu, Maccarone, \& Zepf]{kmz02}
Kundu, A., Maccarome, T. J., \& Zepf, S. E. 2002, ApJ, 574, L5

\bibitem[{{Larsen} {et~al.}(2003){Larsen}, {Brodie}, {Beasley}, {Forbes},
  {Kissler-Patig}, {Kuntschner}, \& {Puzia}}]{LBB+03}
  {Larsen}, S.~S., {Brodie}, J.~P., {Beasley}, M.~A., {Forbes}, D.~A.,
  {Kissler-Patig}, M., {Kuntschner}, H., \& {Puzia}, T.~H. 2003, \apj, 585,
  767

\bibitem[Liu et al.(2001)Liu, van Paradijs, \& van den Heuvel]{lvv01}
Liu, Q. Z., van Paradijs, J., \& van den Heuvel, E. P. J. 2001, A\&A, 368,
1021

\bibitem[Mann \& Whitney(1947)]{mw47}
Mann, H. B., \& Whitney, D. R. 1947, Ann.\ Math.\ Stat., 18, 50

\bibitem[Miller \& Hamilton(2002)]{mh02}
Miller, M. C., \& Hamilton, D. P. 2002, MNRAS, 330, 232

\bibitem[Monet D., et al.(1998)]{Mon+98}
Monet D., et al.\ 1998,
USNO-A V2.0, A Catalog of Astrometric Standards
(Flagstaff: U.S.\ Naval Observatory)

\bibitem[Portegies Zwart \& McMillan(2000)]{pzm00}
Portegies Zwart, S. F., \& McMillan, S. L. 2000, ApJ, 528, L17

\bibitem[Primini et al.(1993)Primini, Forman, \& Jones]{pfj93}
Primini, F. A., Forman, W., \& Jones, C. 1993, ApJ, 410, 615

\bibitem[{{Puzia} {et~al.}(2002){Puzia}, {Zepf}, {Kissler-Patig}, {Hilker},
 {Minniti}, \& {Goudfrooij}}]{PZK+02}
{Puzia}, T.~H., {Zepf}, S.~E., {Kissler-Patig}, M., {Hilker}, M.,
{Minniti},
  D., \& {Goudfrooij}, P. 2002, \aap, 391, 453

\bibitem[Randall et al.(2003)Randall, Sarazin, \& Irwin]{rsi03}
Randall, S. W., Sarazin, C. L., \& Irwin, J. A. 2003, ApJ, submitted (RSI)

\bibitem[Ritter(1999)]{rit99}
Ritter, H. 1999, MNRAS, 309, 360

\bibitem[Sarazin et al.(2000)Sarazin, Irwin, \& Bregman]{sib00}
Sarazin, C. L., Irwin, J. A., \& Bregman, J. N. 2000, ApJ, 544, L101 (SIB)

\bibitem[Sarazin et al.(2001)Sarazin, Irwin, \& Bregman]{sib01}
Sarazin, C. L., Irwin, J. A., \& Bregman, J. N. 2001, ApJ, 556, 533 (SIB)

\bibitem[Sarazin et al.(2003)Sarazin, Angelini, \& Sivakoff]{SAS03}
Sarazin, C.~L., Angelini, L., \& Sivakoff, G.~R. 2003, ApJ, submitted

\bibitem[Sivakoff et al.(2003)Sivakoff, Sarazin, \& Irwin]{ssi03}
Sivakoff, G. R., Sarazin, C. L., \& Irwin, J. A.  2003, ApJ, submitted

\bibitem[Supper et al.(1997)]{shp+97}
Supper, R., Hasinger, G., Pietsch, W., Tr\"umper, J., Jain, A., Magnier, E.
A., Lewin, W. H. G., \& van Paradijs, J. 1997, A\&A, 317, 328

\bibitem[Swartz et al.(2002)]{sgs+02}
Swartz, D. A., Ghosh, K. K., Suleimanov, V., Tennant, A. F., \& Wu, K.
2002, ApJ, 574, 382

\bibitem[Tonry et al.(2001)]{tdb+01}
Tonry, J. L., Dressler, A., Blakeslee, J. P., Ajhar, E. A.,
Fletcher, A. B., Luppino, G. A., Metzger, M. R., \& Moore, C. B.
2001, ApJ, 546, 681

\bibitem[Verbunt \& van den Heuvel(1995)]{vvdh95}
Verbunt, F., \& van den Heuvel, E. P. J. 1995, in
X-ray Binaries,
ed.\ W. Lewin, J. van Paradijs, \& E. van den Heuvel (Cambridge: Cambridge
Univ.\ Press), 457

\bibitem[White \& Angelini(2001)]{wa01}
White, N. E., \& Angelini, L. 2001, ApJ, 561, L101

\bibitem[White \& Ghosh(1998)]{wg98}
White, N. E., \& Ghosh, P. 1998, ApJ, 504, L31

\bibitem[White et al.(1995)White, Nagase, \& Parmar]{wnp95}
White, N. E., Nagase, F., \& Parmar, A. N. 1995, in X-ray Binaries,
ed.\ W. Lewin, J. van Paradijs, \& E. van den Heuvel (Cambridge: Cambridge
Univ.\ Press), 1

\bibitem[White et al.(2002)White, Sarazin, \& Kulkarni]{wsk02}
White, R. E. III, Sarazin, C. L., \& Kulkarni, S. R. 2002, ApJ, 571, L23

\bibitem[Wu(2001)]{wu01}
Wu, K. 2001, PASA, 18, 443

\end{thebibliography}
\end{document}